\documentclass[journal,comsoc]{IEEEtran}

\usepackage[T1]{fontenc}

\ifCLASSINFOpdf
  \usepackage[pdftex]{graphicx}
  \graphicspath{{./images/}}
\else
\fi

\usepackage{mathtools}
\usepackage{mathrsfs}
\usepackage{amsmath}
\usepackage{bm}
\usepackage[cmintegrals]{newtxmath}

\ifCLASSOPTIONcompsoc
  \usepackage[caption=false,font=normalsize,labelfont=sf,textfont=sf]{subfig}
\else
  \usepackage[caption=false,font=footnotesize]{subfig}
\fi
\usepackage{cite}
\usepackage{tabularx}
\usepackage{xcolor}
\usepackage{multirow}

\newcommand{\eexpo}[1]{\exp \left(  #1 \right) }
\newcommand{\esine}[1]{\sin \left(  #1 \right) }
\newcommand{\ecosi}[1]{\cos \left(  #1 \right) }

\newcommand{\eabsn}[2]{\left|  #1 \right| ^{#2}}
\newcommand{\eherm}[1]{\left(  #1 \right)^{H}}
\newcommand{\econj}[1]{\left( #1 \right)^{*}}

\newcommand{\eelem}[3]{\left[  #1 \right]^{#2}_{#3} }

\newcommand{\ediag}[1]{\text{diag}\left( #1 \right)}

\newcommand{\enormn}[3]{\left| \left| #1\right| \right| ^{#2}_{#3}}

\newcommand{\ecsize}[2]{\in\mathbb{C}^{#1\times#2}}

\newcommand{\egausdr}[2]{\mathcal{N}\left( #1,#2\right) }
\newcommand{\egausd}[2]{\mathcal{CN}\left( #1,#2\right) }
\newcommand{\eexpod}[1]{\text{Exp}\left( #1\right) }
\newcommand{\egammd}[2]{\Gamma\left( #1,#2\right) }
\newcommand{\eunifc}[2]{\mathcal{U}\left[  #1,#2 \right]  }

\newcommand{\eexpabstwo}[1]{\mathbb{E} \left\lbrace  \left| #1 \right|^2 \right\rbrace}
\newcommand{\eexpabsn}[2]{\mathbb{E} \left\lbrace  \left| #1 \right|^{#2} \right\rbrace}

\newcommand{\eexp}[1]{\mathbb{E} \left\lbrace #1 \right\rbrace}
\newcommand{\evar}[1]{\text{Var} \left\lbrace #1 \right\rbrace}

\newcommand{\ereal}[1]{\Re \left\lbrace #1 \right\rbrace}

\newcommand{\ecomplexity}[1]{\mathcal{O}\left( #1\right) }

\def\mydate{\leavevmode\hbox{\the\year/\twodigits\month/\twodigits\day}}
\def\twodigits#1{\ifnum#1<10 0\fi\the#1}

\newcommand{\eAddFig}[4]{
	\begin{figure}[!t]
		\centering
		\includegraphics[width=#2\linewidth]{#1}
		\vspace{-6mm}
		\caption{#4}
		\label{#3}
	\end{figure}
}

\begin{document}

\title{Non-Coherent Modulation for RIS-Empowered Multi-Antenna OFDM Communications}

\author{Kun Chen-Hu,~\IEEEmembership{Member,~IEEE,} George C. Alexandropoulos,~\IEEEmembership{Senior~Member,~IEEE,} \\and Ana García Armada,~\IEEEmembership{Senior~Member,~IEEE}
	
\thanks{K. C. Hu and A. García Armada are with the Signal Theory and Communications Department, University Carlos III of Madrid, 28911 Leganés, Spain (e-mails: \{kchen, agarcia\}@tsc.uc3m.es).}

\thanks{G. C. Alexandropoulos is with the Department of Informatics and Telecommunications, National and Kapodistrian University of Athens, 15784 Athens, Greece (e-mail: alexandg@di.uoa.gr).}

\thanks{This work has been partially funded by project TERESA-ADA (TEC2017-90093-C3-2-R) (MINECO/AEI/FEDER, UE).}
}


\maketitle

\begin{abstract}
The Reconfigurable Intelligent Surface (RIS) constitutes one of the prominent technologies for the next 6-th Generation (6G) of wireless communications. It is envisioned to enhance signal coverage in cases where obstacles block the direct communication from Base Stations (BSs), and when high carrier frequencies are used that are sensitive to attenuation losses. In the literature, the exploitation of RISs is exclusively based on traditional coherent demodulation, which necessitates the availability of Channel State Information (CSI). Given the CSI, a multi-antenna BS or a dedicated controller computes the pre/post spatial coders and the RIS configuration. The latter tasks require significant amount of time and resources, which may not be affordable when the channel is time-varying or the CSI is not accurate enough. In this paper, we consider the uplink between a single-antenna user and a multi-antenna BS and present a novel RIS-empowered Orthogonal Frequency Division Multiplexing (OFDM) communication system based on the differential phase shift keying, which is suitable for high noise and/or mobility scenarios. Considering both an idealistic and a realistic channel model, analytical expressions for the Signal-to-Interference and Noise Ratio (SINR) and the Symbol Error Probability (SEP) of the proposed non-coherent RIS-empowered system are presented. Our extensive computer simulation results verify the accuracy of the presented analysis and showcase the proposed system's performance and superiority over coherent demodulation in different mobility and spatial correlation scenarios.
\end{abstract}

\begin{IEEEkeywords}
Channel estimation, differential modulation, mobility, non-coherent system, reconfigurable intelligent surface.
\end{IEEEkeywords}

%
\IEEEpeerreviewmaketitle

\section{Introduction}
\IEEEPARstart{T}{he} evolving technology of Reconfigurable Intelligent Surface (RIS) \cite{Ris01,di2019smart,Ris03,Ris02} is expected to play a significant role in the evolution of mobile communication systems, from the current 5-th Generation (5G) \cite{nr-211} towards the 6-th Generation (6G) \cite{Samsung}. The high frequency bands will be extensively exploited for mobile communications \cite{nr-901}, such as $3.5$ GHz and millimeter waves, in order to take advantage of the huge available bandwidth and provide a fully enhanced Mobile Broadband (eMBB) experience. As a direct consequence, the coverage in these bands will suffer from attenuation loss, and any obstacle may easily block the communication link. RIS-empowered links is an appealing solution to both improve and extend the signal transmitted by either the Base Station (BS) or User Equipment (UE), without excessively increasing the overall cost of the wireless network.

RISs are lightweight and hardware-efficient artificial planar structures of nearly passive reflective elements \cite{Ris02} that enable desired dynamic transformations on the signal propagation environment in wireless communications~\cite{di2019smart}. They can support a wide variety of electromagnetic functionalities \cite{WavePropTCCN}, ranging from perfect and controllable absorption, beam and wavefront shaping to polarization control, broadband pulse delay, radio-coverage extension, and harmonic generation. The RIS technology is envisioned to coat objects in the wireless environment~\cite{di2019smart} (e.g., building facades and room walls), and can operate either as a reconfigurable beyond Snell's law reflector \cite{Ris01}, or as an analog receiver~\cite{hardware2020icassp} or lens~\cite{RISLens2021} when equipped with a single Radio-Frequency (RF) chain, or as a transceiver with multiple relevant RF chains~\cite{DMA2020}.  

The exploitation of RISs is mainly based up to date on the classical Coherent Demodulation Scheme (CDS) \cite{8683663,9053695,8879620,yuanxiaojun_ce,RisChanEst01,RisChanEst02,RisChanEst04,PARAFAC2020,RisChanEst05,RISpatternMod2021}, where the knowledge of Channel State Information (CSI) is essential for the optimized configuration of the RIS tunable elements and the demodulation of the signal at the receiving node. An approach for estimating the cascaded channel matrix \cite{8879620,yuanxiaojun_ce}, which encompasses the joint effect of the signal propagation over the BS-RIS and RIS-UE links, was proposed in \cite{9053695}. Note that this channel cannot be easily decoupled with nearly passive RISs that do not possess receive RF chains. However, the estimation of the cascaded channel depends on the RIS configuration, which implies that the minimum required training periods equals the total number of configurations. Consequently, this estimation overhead becomes prohibitive as the numbers of RIS elements and configurations increase \cite{Ris04}. Late interests in reducing the channel estimation overhead focus on efficient decompositions of the received signal \cite{PARAFAC2020} and on designing RIS configurations tailored for channel estimation \cite{RISpatternMod2021}. In the vast majority of the CDS works, Time Duplex Division (TDD) is typically adopted and the CSI is assumed to be estimated in the uplink and then reused in the downlink. To this end, the coherence time is always considered to be long enough to cope with the channel training and uplink/downlink data transmission stages. Given the CSI availability, the BS computes the best pair of precoder/combiner as well as the set of RIS elements' configuration, which is communicated to the RIS via a side control link. This processing task is not straightforward due to the fact that a non-convex design optimization needs to be solved, increasing the operational complexity of the RIS-empowered communication system. When Orthogonal Frequency Division Multiplexing (OFDM) \cite{ofdm1,ofdm2} is taken into account, the complexity of the channel estimation and optimization scales with the number of subcarriers \cite{RisChanEst04,RisChanEst05}. Several late studies have focused on accelerating this optimization using alternative methods at the expense of sacrificing the performance, such as sub-optimal optimizations \cite{Ris01,Ris03,huang2018energy}, or configuring sets of contiguous passive reflective elements with the same phase value to decrease the number of variables to be optimized \cite{RisChanEst01,RisChanEst02}. 

The Non-CDS (NCDS) is an alternative demodulation scheme that does not require CSI, hence, reducing the undesirable signaling overhead and increasing the effective data rate of the communication system \cite{nc1,nc2,nc3}. This scheme is realized with reduced complexity transmission and reception, which implies cheaper transceiver hardware devices and lower latency for processing. Recently, NCDS has been combined with massive Multiple-Input Multiple-Output (MIMO) systems \cite{Andrea01,Andrea02,Ana2015,Victor2017,Victor2018,Victor2019,Kun2019,Kun2020,Manu2020}, where it was shown to provide a significant performance gain compared to CDS for some 5G challenging scenarios, such as vehicular and low-latency communications. It was highlighted in \cite{Andrea01,Andrea02} that NCDS is more robust than CDS for low Signal-to-Noise Ratio (SNR) scenarios, where the latter scheme additionally suffers from noise pollution in the channel estimates. In \cite{Ana2015,Victor2017,Victor2018,Victor2019,Kun2019,Kun2020,Manu2020}, the use of differential Phase Shift Keying (PSK) \cite{Adachi1996} was proposed. Those works showcased that this modulation scheme is robust in very fast time-varying channels, because it only requires that the channel response is quasi-static over two contiguous symbols. Additionally, \cite{Kun2019,Kun2020,Manu2020} proved the superiority of NCDS over CDS in terms of throughput, due to the fact that reference signals designed for channel tracking can be fully avoided.

To the best knowledge of the authors, RIS-empowered wireless communications based on NCDS have not been proposed yet. Similar to the case of massive MIMO, communication systems including RISs with large numbers of passive elements require significant numbers of reference signals to obtain the estimation of the cascaded channels, which can be avoided when using NCDS. Motivated by these facts, in this paper we propose a novel combination of the latter technologies targeting new broadband applications of 5G and 6G systems, such as long-range communications (low SNR cases), vehicular communications (mobility scenarios), and low-latency communications. The main contributions of the paper are summarized as follows:
\begin{itemize}
	\item We present a RIS-empowered Single-Input Multiple-Output (SIMO) OFDM system with differential PSK modulation. This combination requires neither channel estimation nor solving a non-convex optimization problem. Consequently, the channel training stage is no longer required and the side link to control the RIS is removed, since the passive elements of the RIS can be configured by any random phases. Hence, the proposed solution is not only able to improve the efficiency of the system, but it is also capable of reducing the processing complexity, especially for broadband multi-carrier waveforms, simplifying the massive deployment of RISs.
	\item The Signal-to-Interference plus Noise Ratio (SINR), determining the useful signal over the self-interference and thermal noise terms, of the proposed RIS-empowered NCDS system is analytically characterized over both an Independent and Identically Distributed (IID) Rayleigh channel model and a realistic geometric wideband channel model \cite{chan_meas1,chan_meas2,chan_meas3}, including UE mobility as well as temporal and spatial correlations.
	\item Capitalizing on the approach of \cite{Manu2020}, we derive approximate analytical expressions for the Symbol Error Probability (SEP) of the proposed system, when operating under any of the two considered channel models. 	
	\item Our simulation results verify the accuracy of the presented analysis and highlight the superiority of the proposed NCDS system over a relevant CDS one. The inefficiency of the cascaded channel estimation in the CDS system is numerically assessed using the 5G numerology, which is an additional figure of merit to show its weakness as compared to the proposed NCDS approach. Besides, the performance of the proposed NCDS does not suffer from any performance penalization when low resolution quantization (even at $1$ bit) is considered for the RIS phase configurations, unlike CDS.
\end{itemize}

The remainder of the paper is organized as follows. Section~\ref{sec:system_model} introduces the system model and the two considered channel models, while Section~\ref{sec:baseline} provides a summary of the CDS when applied in a RIS-empowered link. Section \ref{sec:nc-diff} details the implementation of the proposed differential PSK scheme and presents the analytical expressions for the SINR. Section~\ref{sec:sercomplex} includes the approximate SEP analysis and Section~ \ref{sec:num_res} discusses the performance assessment results. Finally, Section~\ref{sec:conclusion} concludes the paper.

\textbf{Notation:} Matrices, vectors, and scalar quantities are denoted by boldface uppercase, boldface lowercase, and normal letters, respectively. $\left[\mathbf{A}\right]_{mn}$ denotes the element in the $m$-th row and $n$-th column of $\mathbf{A}$, $\left[\mathbf{A}\right]_{:,n}$ is $\mathbf{A}$'s $n$-th column, $\mathbf{v}_{\rm max}(\mathbf{A})$ is $\mathbf{A}$'s principal eigenvector, and $\left[\mathbf{a}\right]_{n}$ represents the $n$-th element of $\mathbf{a}$. $\ediag{\mathbf{a}}$ denotes a diagonal matrix whose diagonal elements are formed by $\mathbf{a}$'s elements. $\Re(\cdotp)$ and $\Im(\cdotp)$ represent the real and imaginary part of a complex number, respectively, and $\jmath$ is the imaginary unit, while $*$ denotes the convolution operation. $\enormn{\cdotp}{2}{F}$ denotes the squared Frobenius norm. $\eabsn{\cdotp}{}$ is the absolute value. $\mathbb{E}\left\lbrace \cdotp \right\rbrace$ represents the expected value of a random variable, $\text{Var}\left\lbrace \cdotp \right\rbrace$ denotes the variance, and $\mathcal{CN}(0,\sigma^2)$ represents the circularly-symmetric and zero-mean complex normal distribution with variance $\sigma^2$. $\eexpod{\lambda}$ accounts for the Exponential distribution with rate parameter $\lambda$, $\egammd{k}{\theta}$ is the Gamma distribution with shape parameter $k$ and scale parameter $\theta$, such that the mean $\mu = k\theta$ and the variance $\sigma^{2} = k\theta^2$. $\eunifc{a}{b}$ denotes the continuous uniform distribution with the minimum value $a$ and maximum value $b$.



\section{System and Channel Models}
\label{sec:system_model}
This section describes the considered mobile communication link empowered by an RIS. In addition, an idealistic and a realistic propagation channel models are detailed.

\subsection{Considered Mobile Communication Scenario}
The considered mobile communication scenario comprises a BS, an RIS, and a single-antenna UE (see Fig. \ref{fig:scenario}). The BS is equipped with a uniform rectangular array (URA) consisting of $B=B_{H}B_{V}$ antenna elements, where $B_{H}$ and $B_{V}$ denote the number of elements in the horizontal and vertical axes, respectively, and the distance between any two contiguous elements in their respective axes is given by $d_{H}^{\text{BS}}$ and $d_{V}^{\text{BS}}$. Analogously to the BS, the RIS is built by $M=M_{H}M_{V}$ fully passive reflecting unit elements, whose respective distances between elements are given by $d_{H}^{\text{RIS}}$ and $d_{V}^{\text{RIS}}$. The UE is constrained to have a single antenna element. Throughout the paper, different numbers $B$ of antennas at the BS will be considered; a small value for $B$ corresponds to a small and low-complexity BS, while a large $B$ indicates that the BS is equipped with a massive MIMO array. On the other hand, the number $M$ of passive elements at the RIS may be extremely large due its low fabrication and operation costs.

\eAddFig{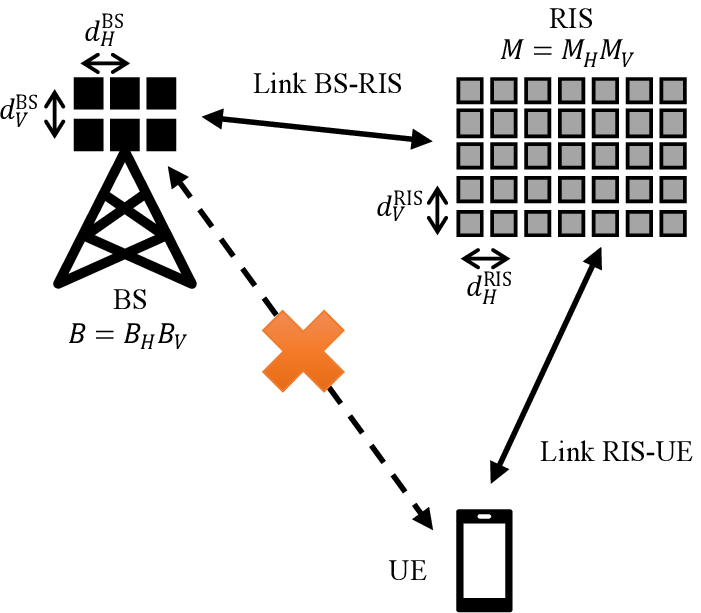}{1}{fig:scenario}{The RIS-empowered wireless communication link comprising a multi-antenna BS, a multi-element passive RIS, and a single-antenna mobile UE.}

Regarding the signal propagation, it is assumed that a direct communication link between the BS and UE (BS-UE) is absent, due to the presence of blockages. Therefore, the communication between the BS and UE must be established through the RIS, via the BS-RIS and RIS-UE communication links. This work focuses on the uplink case, where the UE transmits both reference (if needed, e.g., in CDS) and data symbols to the BS through the RIS. It is understood that other UEs may be multiplexed in different orthogonal (time or frequency) resources; this extension is left for future work. It is assumed that, at each communication frame, the UE transmits a frame of $N$ contiguous OFDM symbols of $K$ subcarriers each. In order to avoid the Inter-Symbol and Inter-Carrier Interferences (ISI and ICI), the length $L_{CP}$ of the cyclic prefix must be long enough to absorb the effective multipath produced by the cascaded channel, namely the sum of the lengths of each of the channel responses of both BS-RIS and RIS-UE channels. The baseband representation of the received signal $\mathbf{y}_{k,n}\ecsize{B}{1}$ at the BS in the $k$-th subcarrier, with $1\leq k \leq K$, and $n$-th OFDM symbol, with $1 \leq n \leq N$, is given by
\begin{equation}\label{eqn:model_y}
\mathbf{y}_{k,n} = \mathbf{q}_{k,n}x_{k,n}  + \mathbf{v}_{k,n}, 
\end{equation}
where $x_{k,n}\in\mathbb{C}$ denotes the symbol transmitted from the UE at the $k$-th subcarrier and $n$-th OFDM symbol, whose transmit power is $\eexpabstwo{x}=P_{x}$, $\mathbf{v}_{k,n}\ecsize{B}{1}$ represents the Additive White Gaussian Noise (AWGN) vector which is distributed as $\eelem{\mathbf{v}_{k,n}}{}{b}\sim\egausd{0}{\sigma_{v}^{2}}$, and $\mathbf{q}_{k,n}\ecsize{B}{1}$ is the effective RIS-empowered cascaded channel frequency response, which can be decomposed for $1\leq k \leq K$ and $1 \leq n \leq N$ as
\begin{equation}\label{eqn:model_chan1}
\mathbf{q}_{k,n}\triangleq\mathbf{H}_{k,n}\mathbf{\Psi}_{n}\mathbf{g}_{k,n}=\sum_{m=1}^{M}\eelem{\boldsymbol{\psi}_{n}}{}{m} \eelem{\mathbf{H}_{k,n}}{}{:,m}\eelem{\mathbf{g}_{k,n}}{}{m},
\end{equation}
where $\mathbf{H}_{k,n}\ecsize{B}{M}$ is the channel frequency response matrix between BS and RIS, $\mathbf{g}_{k,n}\ecsize{M}{1}$ accounts for the channel frequency response vector between RIS and the single UE of interest, and $\mathbf{\Psi}_{n}\triangleq\ediag{\boldsymbol{\psi}_n}\in\mathbb{C}^{M\times M}$ is a diagonal matrix accounting for the effective phase configurations applied by the passive reflecting elements of the RIS at the $n$-th OFDM symbol, where $\boldsymbol{\psi}_n\in\mathbb{C}^{M\times M}$ is defined as
\begin{equation}\label{eqn:panel}
	\boldsymbol{\psi}_n \triangleq \begin{bmatrix} \eexpo{\jmath \psi_{n,1}} & \cdots & \eexpo{\jmath\psi_{n,M}} \end{bmatrix}, 
\end{equation}
with $\psi_{n,m}$ for $1\leq m \leq M$ representing the phase shift of the $m$-th passive element of the RIS panel.

\subsection{Independent and Identically Distributed (IID) Channel Model}
As a benchmark, we consider the case where the elements of the channel frequency response for both links (BS-RIS and RIS-UE) are IID. This channel model will be used for the purpose of upper-bounding the performance of the proposed NCDS, and comparing it with a more realistic geometric wideband channel model. In this case, the propagation channels (BS-RIS and RIS-UE) at the $k$-th subcarrier and $n$-th OFDM symbol are modeled as
\label{subsec:iid}
\begin{equation}\label{eqn:chan_h_iid}
	\mathbf{H}_{k,n} \triangleq \sqrt{L_{\alpha}} \mathbf{A}_{k,n}, \quad \eelem{\mathbf{A}_{k,n}}{}{bm} \sim \egausd{0}{\sigma_{\alpha}^{2}},
\end{equation}
\begin{equation}\label{eqn:chan_g_iid}
	\mathbf{g}_{k,n} \triangleq \sqrt{L_{\beta}} \mathbf{b}_{k,n}, \quad \eelem{\mathbf{b}_{k,n}}{}{m} \sim \egausd{0}{\sigma_{\beta}^{2}},
\end{equation}
\begin{equation*}
	1\leq b \leq B, \quad 1 \leq m \leq M, 
\end{equation*}
where $L_{\alpha}$ and $L_{\beta}$ denote the large-scale gains of the BS-RIS and RIS-UE links, respectively, and $\mathbf{A}_{k,n}\ecsize{B}{M}$ and $\mathbf{b}_{k,n}\ecsize{M}{1}$ model the small-scale fading for their respective channels, according to a Rayleigh distribution. Hence, the average gain of each link is $\sigma_{h}^{2} = L_{\alpha}\sigma_{\alpha}^{2}$ and $\sigma_{g}^{2} = L_{\beta}\sigma_{\beta}^{2}$, respectively.

Moreover, it is assumed that the channel between BS and RIS remains quasi-static, while the channel between the RIS and UE may suffer from time variability given as
\begin{equation} \label{eq:dopplerchan}
\eexp{\econj{\eelem{\mathbf{g}_{k,n}}{}{m}}\eelem{\mathbf{g}_{k,n'}}{}{m}}
=\eabsn{J_{0}\left(2\pi f_{d}  \frac{\Delta n}{\Delta f}\left( 1+\frac{L_{CP}}{K}\right) \right)}{},
\end{equation}
\begin{equation*}
\Delta n = n'-n, \quad 1 \leq k \leq K, \quad 1 \leq n \leq N, \quad 1 \leq m \leq M,
\end{equation*}
where $J_{0}\left( \cdot \right) $ denotes the zero-th order Bessel function of the first kind \cite{bessel}, and $f_{d}$ and $\Delta f$ represent the Doppler frequency shift experienced by the signal transmitted from the UE and the distance between two contiguous subcarriers, respectively, both measured in Hz.

\subsection{Geometric Wideband Channel Model}
\label{subsec:geo}
For a more realistic performance evaluation, the two links (BS-RIS and RIS-UE) are characterized with a geometric wideband model \cite{chan_meas1,chan_meas2,chan_meas3}, made up of the superposition of several separate clusters, where each of them has a different value of delay and gain. Moreover, each cluster is comprised of a certain number of rays with different angles of arrival and departure. The delays and geometrical positions of each cluster/ray are typically characterized by the Delay and Angular Spreads (DS and AS), respectively. Each propagation model has its own definition for the value of these parameters, the number of clusters/rays and how the delays and rays are distributed for a given propagation environment. Then, given the information of these clusters/rays, the array steering vectors of both transmitter and receiver are included to model the spatial correlation due to the array responses. Therefore, note that this channel model is able to account for the spatial correlation considering both the given antenna array response of the BS and RIS, as well as the geometrical positions of all clusters/rays. 

In order to provide a realistic evaluation of the system, the propagation channel model recommended for 5G \cite{nr-901} is chosen, where it is assumed that all clusters have the same number of rays and all the rays of a particular cluster have the same delay and gain. The power-delay profile follows an exponential distribution whose standard deviation is the DS; the azimuth angles of arrival/departure are modeled by a wrapped Gaussian distribution which is characterized by the Azimuth angular Spread of Arrival and Departure (ASA and ASD); and the zenith angles of arrival/departure are modeled by a Laplacian distribution, also characterized by are the Zenith angular Spread of Arrival and Departure (ZSA and ZSD).

The channel response between the BS and RIS at the $k$-th subcarrier and $n$-th OFDM symbol can be described as
\begin{equation}\label{eqn:chan_h}
\begin{split}
&\mathbf{H}_{k,n} \triangleq 
\sqrt{L_{\alpha}}
\sum_{c=1}^{C_{\alpha}}\sum_{r=1}^{R_{\alpha}}
\alpha_{n}^{cr} \mathbf{a}_{\rm BS}\left(\phi_{n}^{cr},\theta_{n}^{cr}\right)
\mathbf{a}_{\rm RIS}^{H}\left(\bar{\varphi}_{n}^{cr},\bar{\vartheta}_{n}^{cr}\right) \times \\
& \times \eexpo{-j\frac{2\pi}{K}\left( k-1\right) {\tau_{\alpha}}_{n}^{c}}, \quad \alpha_{n}^{cr}\sim\egausd{0}{\frac{\sigma_{\alpha,c}^{2}}{R_{\alpha}}},
\end{split}
\end{equation}
where $C_{\alpha}$ is the number of clusters, $R_{\alpha}$ represents the number of rays for each cluster, ${\tau_{\alpha}}_{n}^{c}$ accounts for the delay of the $c$-th cluster measured in samples, $\alpha_{n}^{cr}$ is the channel coefficient for the $c$-th cluster and $r$-th ray, $\sigma_{\alpha,c}^{2}$ is the average gain of the $c$-th cluster, and $\mathbf{a}_{\rm BS}\left(\phi_{n}^{cr},\theta_{n}^{cr}\right)$ accounts for the array steering vector at the BS, and its arguments are the azimuth and elevation angles of arrival, respectively, for the $c$-th cluster and $r$-th ray. The steering vector for the BS is given by
\begin{equation}\label{eqn:ura}
\begin{split}
& \eelem{\mathbf{a}_{BS}\left( \phi, \theta \right) }{}{B_{H}(b_{V}-1)+b_{H}} = \\ 
& \eexpo{\jmath\frac{2\pi}{\lambda} (b_{H}-1) d_{H}^{\text{BS}}\esine{\theta}\ecosi{\phi} } \times\\
& \eexpo{\jmath\frac{2\pi}{\lambda} (b_{V}-1) d_{V}^{\text{BS}}\esine{\theta}\esine{\phi} }, \\
& 1 \leq b_{H} \leq B_{H}, \quad 1 \leq b_{V} \leq B_{V},
\end{split}
\end{equation}
where $\lambda$ is the wavelength, 
Similarly to the BS, $\mathbf{a}_{\rm RIS}\left(\bar{\varphi}_{n}^{cr},\bar{\vartheta}_{n}^{cr}\right)$ denotes the steering vector for the RIS, and its arguments are the azimuth and elevation angles of departure, respectively, for the $c$-th cluster and $r$-th ray. The expression for the steering vector is the same as described in (\ref{eqn:ura}), replacing respectively the set ($B_{H}$, $B_{V}$, $d_{H}^{\text{BS}}$, $d_{V}^{\text{BS}}$) by the set ($M_{H}$, $M_{V}$, $d_{H}^{\text{RIS}}$, $d_{V}^{\text{RIS}}$).

The channel response between the RIS and UE at the $k$-th subcarrier and $n$-th OFDM symbol is given by
\begin{align}\label{eqn:chan_g}
\mathbf{g}_{k,n} \triangleq&
\sqrt{L_{\beta}}
\sum_{c=1}^{C_{\beta}}\sum_{r=1}^{R_{\beta}}
{\beta}_{n}^{cr} 
\mathbf{a}_{\rm RIS}\left(\varphi_{n}^{cr},\vartheta_{n}^{cr}\right)
\eexpo{-j\frac{2\pi}{K}\left( k-1\right) {\tau_{\beta}}_{n}^{c}},\nonumber \\
& \beta_{n}^{cr}\sim\egausd{0}{\frac{\sigma_{\beta,c}^{2}}{R_{\beta}}},
\end{align}
where $C_{\beta}$ is the number of clusters, $R_{\beta}$ represents the number of rays for each cluster, ${\tau_{\beta}}_{n}^{c}$ accounts for the delay of the $c$-th cluster measured in samples, $\alpha_{n}^{cr}$ is the channel coefficient for the $c$-th cluster and $r$-th ray, $\sigma_{\beta,c}^{2}$ is the average gain of the $c$-th cluster, $\mathbf{a}_{\rm RIS}\left(\varphi_{n}^{cr},\vartheta_{n}^{cr}\right)$ accounts for the array steering vector at the RIS, and its arguments are the azimuth and elevation angles of arrival, respectively, for the $c$-th cluster and $r$-th ray. Hence, similarly to the IID channel model, the average gain of each link can be defined as
\begin{equation}\label{eqn:chan_gain}
\sigma_{h}^{2} = L_{\alpha}\sum_{c=1}^{C_{\alpha}}\sigma_{\alpha,c}^{2}, 
\quad \sigma_{g}^{2} = L_{\beta}\sum_{c=1}^{C_{\beta}}\sigma_{\beta,c}^{2}.
\end{equation}
Similarly to the IID channel model case, the link between RIS and the UE of interest may suffer from a Doppler shift due to the mobility also charaterized by (\ref{eq:dopplerchan}), while the link between BS and RIS is assumed to be quasi-static.

\section{Baseline RIS-Empowered System Based on CDS}
\label{sec:baseline}
In this section, we detail the operation of the considered baseline RIS-empowered system that is based on conventional CDS. The performance of this reference system will be compared with that based on the proposed NCDS, which will be presented in the following section. According to \cite{RisChanEst01,RisChanEst02,8879620,RisChanEst04}, in order to be able to fully exploit the benefits of the RIS, the system requires to perform a channel training stage before the data transmission stage. This channel training stage mainly consists of three tasks: cascaded channel sounding, processing for the desired RIS configuration, and parameter feedback (see Fig. \ref{fig:ris_chan_est}). These tasks are detailed in the following subsections.

\subsection{Cascaded Channel Sounding}
Let us first rewrite the RIS-empowered cascaded channel frequency response, defined in (\ref{eqn:model_chan1}) for each $k$-th subcarrier and each $n$-th OFDM symbol, as follows:
\begin{equation}\label{eqn:cascaded_chan}
\begin{split}
\mathbf{q}_{k,n} &= \sum_{m=1}^{M}\eelem{\boldsymbol{\psi}_{n}}{}{m} \mathbf{c}_{k,n,m}
\\&=\underbrace{\left[\mathbf{c}_{k,n,1}\,\mathbf{c}_{k,n,2}\,\cdots\,\mathbf{c}_{k,n,m}\right]}_{\triangleq\mathbf{C}_{k,n,M}}\boldsymbol{\psi}_{n},   
\end{split}
\end{equation}
where the $B$-element vector $\mathbf{c}_{k,n,m}\triangleq\eelem{\mathbf{H}_{k,n}}{}{:,m}\eelem{\mathbf{g}_{k,n}}{}{m}$ and the $B\times M$ matrix $\mathbf{C}_{k,n,m}$ include the cascaded channel produced by the BS-RIS and RIS-UE channels. Inspecting (\ref{eqn:cascaded_chan}), it becomes apparent that the BS needs to estimate in the uplink the cascaded channel for each $m$-th passive element of the RIS for all $K$ subcarriers at each coherence time. In order to accomplish this channel estimation per coherence channel time, the passive elements of the RIS are firstly configured with the known angle values $\tilde{\psi}_{n,m}$ $\forall$$n,m$ and then the UE transmits the known reference signal $x_{k,n}=p_{k,n}$. The BS finally estimates the cascaded channel via a typical estimation method. The case of the Least-Squares (LS) \cite{ls} method yields the following estimate: \begin{equation}\label{eqn:est_chan}
	\widehat{\mathbf{c}}_{k,n,m} = \mathbf{c}_{k,n,m} + \mathbf{v}_{k,n} \frac{1}{p_{k,n}\eexpo{\jmath\tilde{\psi}_{n,m}}}\,\,\,\forall k,n,m,
\end{equation}
The latter cascaded channel sounding procedure requires at least $M$ symbol periods to collect all the estimates, one per each passive element of the RIS. The bottom of Fig.~\ref{fig:ris_chan_est} shows an example of the cascaded channel estimation for two particular passive elements of the RIS in two time instants (red and blue arrows). Note that the passive elements of the RIS may not be properly configured during this sounding task ($\tilde{\psi}_{n,m}$, $1\leq m \leq M$), and the RIS may not be able to adequately reflect the transmitted reference signals from the UE towards the BS. This implies, in practice, that more than $M$ symbol periods will be required to use an acceptable RIS configuration during channel sounding. Several works in the literature proposed techniques to improve the quality of channel estimation \cite{RisChanEst01,RisChanEst02,8879620,RisChanEst04}, however, none of them is able to significantly reduce the total number of required training symbol periods. Hence, an RIS equipped with a very large number of reflective elements is not recommendable due to the fact that the length of the training period scales with this value, shrinking the time devoted for data transmission at each channel coherence time \cite{Ris04}.

\eAddFig{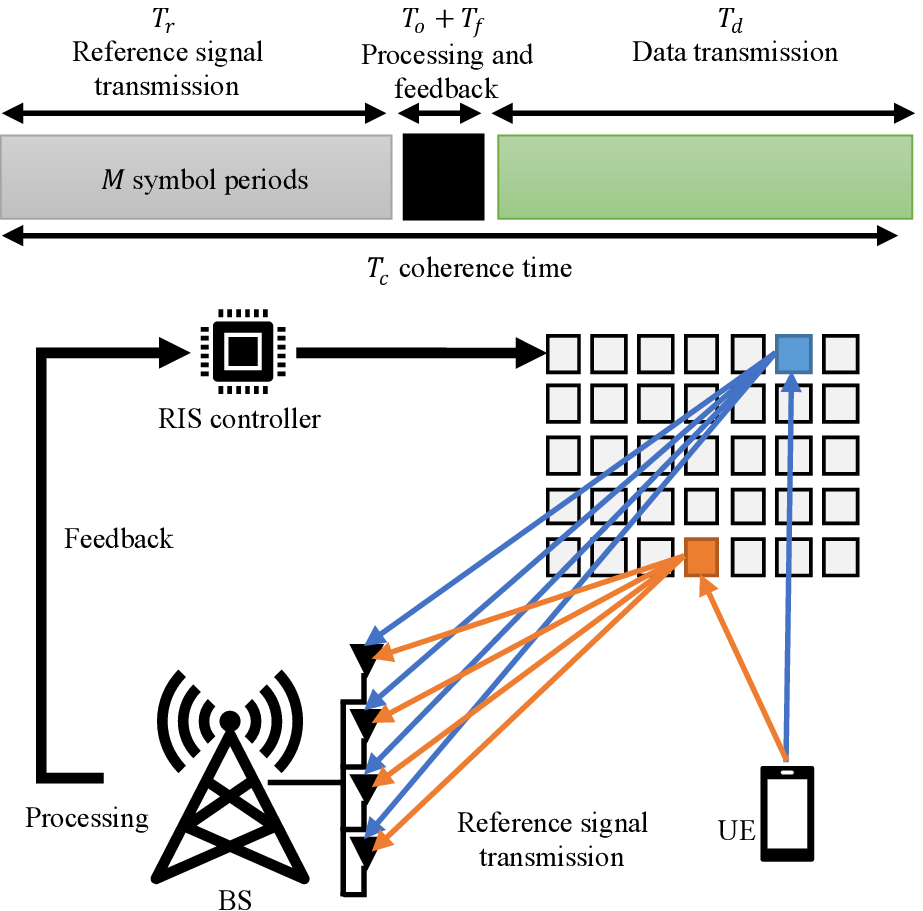}{1}{fig:ris_chan_est}{Channel estimation procedure for the baseline RIS-empowered system based on CDS.}

\subsection{RIS Configuration Optimization and Feedback}
Taking into account the received signal model in (\ref{eqn:model_y}) and the notation in (\ref{eqn:cascaded_chan}), the post-processed symbol for the $k$-subcarrier and $n$-th OFDM symbol at the BS is given by
\begin{equation}\label{eqn:co_rx}
\begin{split}
z_{k,n} & \triangleq \mathbf{w}_{k,n}^H \mathbf{y}_{k,n} = 
\mathbf{w}_{k,n}^H \mathbf{q}_{k,n} x_{k,n} + \mathbf{w}_{k,n}^H \mathbf{v}_{k,n} = \\
& = \mathbf{w}_{k,n}^H\mathbf{C}_{k,n,m}\boldsymbol{\psi}_{n}x_{k,n} + \mathbf{w}_{k,n}^H\mathbf{v}_{k,n}, 
\end{split}
\end{equation}
where $\mathbf{w}_{k,n}\ecsize{B}{1}$ represents the BS combiner vector at the $k$-subcarrier and $n$-th OFDM symbol. 

Once the cascaded channel estimates are available at the BS, the best pair of combiner vector and RIS phase configurations (i.e., $\mathbf{w}_{k,n}$ and $\mathbf{\Psi}_{n}$), capable of increasing the quality of the link and enhancing the end-to-end link's data rate, need to be obtained. Then, the computed RIS phase configurations must be sent to the RIS though a side link, and therefore, the RIS can be reconfigured accordingly. Given the cascaded channel estimate $\eelem{\widehat{\mathbf{c}}_{k,n}}{}{m}$, the instantaneous estimated SNR after post-coding for the CDS is obtained as
\begin{equation}\label{eqn:co_snr}
\rho_{c}(\mathbf{w}_{k,n}, \mathbf{\Psi}_{n}) = \frac{P_{x}}{\sigma_{v}^{2}}\frac{\eabsn{\mathbf{w}_{k,n}^H \mathbf{C}_{k,n,m}\boldsymbol{\psi}_{n}}{2}}{\|\mathbf{w}_{k,n}\|_2^2}.
\end{equation}
Given (\ref{eqn:co_snr}), a proper choice of the pair receive combiner $\mathbf{w}_{k,n}$ and RIS phase configuration $\mathbf{\Psi}_{n}$ $\forall$$k,n$ is capable of increasing the conceived instantaneous SNR, $\rho_{c}$, at BS. This pair can be obtained as the solution of the following optimization problem:
\begin{equation} \label{eq:maxrate}
\begin{split}
\underset{ \mathbf{w}_{k,n}, \mathbf{\Psi}_{n} }{\text{max}} 
& \quad  \rho_{c}(\mathbf{w}_{k,n}, \mathbf{\Psi}_{n})\\ 
\text{s.t.}& \quad \|\mathbf{w}_{k,n}\|_2^2\leq1, \quad  0 \leq \psi_{n,m} \leq 2\pi,\,\,\,\forall k,n,m.
\end{split}
\end{equation}
Similar to \cite{Ris01,Ris02,Ris03}, this optimization problem is not jointly convex in terms of the unknowns $\mathbf{w}_{k,n}$ and $\mathbf{\Psi}_{n}$, and in order to solve it optimally, significant amount of time and computation resources are needed. It is also noted that the complexity of this optimization problem is significantly increased when the number of variables to be optimized (i.e., $B$ variables in $\mathbf{w}_{k,n}$ and $M$ variables for $\mathbf{\Psi}_{n}$) is large. Furthermore, it is even more complex when OFDM is used \cite{RisChanEst04,RisChanEst05}, due to the fact that the combiner must be computed for each $k$-th subcarrier out of the $K$ total. Consequently, a CDS optimized as above is not compatible with a RIS built by a large number $M$ of passive reflective elements \cite{Ris04}, and the problem is aggravated for broadband transmission in frequency-selective channels. Alternatively, at the expense of sacrificing performance and accelerating the solution, (\ref{eq:maxrate}) can be sub-optimally tackled \cite{Ris06}
with the following iterative approach. Keeping $\boldsymbol{\psi}_{n}$ fixed $\forall$$n$, the optimum $\mathbf{w}_{k,n}$ is obtained $\forall$$k$ as $\mathbf{w}_{k,n}^{\rm opt} = \mathbf{v}_{\rm max}(\mathbf{C}_{k,n,m} \boldsymbol{\psi}_{n}\boldsymbol{\psi}_{n}^H \mathbf{C}_{k,n,m}^H)$, while for fixed $\mathbf{w}_{k,n}$ $\forall$$k,n$, the optimum $\boldsymbol{\psi}_{n}$ is computed as $\boldsymbol{\psi}_{n}^{\rm opt} = \exp{(\jmath \arg(\mathbf{C}_{k,n,m}^H \mathbf{w}_{k,n}))}$. This alternating optimization procedure continues until convergence or until the maximum affordable number of iterations is reached. While many previous works assumed perfect channel estimation when the noise is high or even moderate, $\mathbf{w}_{k,n}$ and $\mathbf{\Psi}_{n}$ computed from the latter algorithm with noisy channel estimations will be far from their counterparts maximizing the actual instantaneous SNR, hence, the overall link will be significantly degraded.

\subsection{Efficiency Factor}
In most of the previous works \cite{Ris01,Ris02,Ris03,Ris04,Ris05,Ris06}, it has been assumed for the purpose of CDS for the RIS-empowered link that the coherence time, $T_{c}$, is always long enough so that the duration of channel training does not penalize the duration of the data transmission stage. However, even with low mobility, the cascaded channel will suffer from a certain time variability and its estimation must be periodically updated. Consequently, the combiner $\mathbf{w}_{k,n}$ and RIS phase configuration $\mathbf{\Psi}_{n}$ $\forall$$k,n$ need to be updated accordingly. It is hence important to take into account the additional inefficiency produced by the channel training and estimation stages, if a feasible RIS-empowered link is to be designed.

According to \cite{Ris04}, the channel training phase produces a reduction of the communication effective data rate, where the time required for reference signal transmission, processing, and feedback should be taken into account, as shown in Fig. \ref{fig:ris_chan_est}. The feedback time, $T_{f}$, is typically assumed to be negligible. The processing time, $T_{p}$, is difficult to quantify due to the fact that the time required for solving the design optimization problem depends on the chosen numerical method and the amount of resources assigned for this task. Hence, the data rate penalty due to channel sounding is lower-bounded by taking only into account the channel sounding time, $T_{r}$. 

Following \cite{rappa}, the coherence time measured in seconds is given by $T_{c} = 0.423 / f_{d}$ 
and the coherence time measured in the number of OFDM symbols, $N_{c}$, can be computed as
\begin{equation}\label{eqn:Nc}
N_{c} = \frac{\Delta f}{f_{d}}\frac{0.423K}{K+L_{CP}}.
\end{equation}
The effective transmitted power at the UE can be defined as $P_{x}^{\text{eff}}\triangleq\frac{P_{x} }{\eta_{c}}$, where $\eta_{c}$ represents the efficiency factor that takes into account the coherence time and the number of RIS passive elements to be sounded. This efficiency factor is given by 
\begin{equation}\label{eqn:effective_power}
\quad \eta_{c} \triangleq 1-\frac{T_{r}}{T_{c}} \leq 1-\frac{M}{N_{c}}.
\end{equation}
According to \cite{Ris04}, the time and/or power resources devoted for performing the cascaded channel sounding are penalizing the overall performance of the system.


\section{Proposed RIS-Empowered System Based on NCDS with Differential Modulation}
\label{sec:nc-diff}
In order avoid the inefficiencies pointed out above, this work proposes to replace the classical CDS by a NCDS based on differential modulation. This approach does not require training-based channel estimation in order to perform the demodulation and decision, as shown in \cite{Ana2015,Victor2017,Victor2018,Victor2019,Kun2019,Kun2020,Manu2020}.

\subsection{Differential Encoding and Decoding}

At the UE, the data symbols are differentially encoded in the time domain before their transmission as:
\begin{equation} \label{eqn:diff_enc}
x_{k,n}  =
\left\{\begin{array}{@{}cl}
s_{k,n},  & n=1 \\
x_{k,n-1} s_{k,n}, & 2\leq n \leq N\\
\end{array}
\right., \quad 1 \leq k \leq K.
\end{equation}
where $s_{k,n}$ denotes the complex symbol to be transmitted at the $k$-th subcarrier and $n$-th OFDM symbol, that belongs to a PSK constellation and its power is normalized (i.e., $\eabsn{s_{k,n}}{2}=1$). Note that the differential modulation only requires a single reference symbol $s_{k,1}$ at the beginning of the burst in order to allow the differential demodulation, which represents a negligible overhead. The differential modulation can be also implemented in the frequency domain with the same performance \cite{Kun2019}. Before data transmission, the power of differential symbols $x_{k}^{n}$ is scaled according to $P_{x}$.

Given (\ref{eqn:model_y}), the BS performs the differential decoding as
\begin{equation} \label{eqn:diff_dec}
z_{k,n} = \frac{1}{MB} \eherm{\mathbf{y}_{n-1}^{k}}\mathbf{y}_{k,n}= \frac{1}{MB}\sum_{i=1}^{4} I_{i}, 
\end{equation}
\begin{equation*}
2 \leq n \leq N, \quad 1 \leq k \leq K,
\end{equation*}
\begin{equation} \label{eqn:diff_dec1}
I_{1} = \eherm{\mathbf{q}_{n-1}^{k}}\mathbf{q}_{k,n} s_{k,n}, \quad I_{2} = \eherm{\mathbf{q}_{n-1}^{k}x_{n-1}^{k}}\mathbf{v}_{k,n},
\end{equation}
\begin{equation} \label{eqn:diff_dec3}
I_{3} = \eherm{\mathbf{v}_{n-1}^{k}}\mathbf{q}_{k,n} x_{k,n}, \quad I_{4} = \eherm{\mathbf{v}_{n-1}^{k}} \mathbf{v}_{k,n},
\end{equation}
where $I_{1}$ includes the useful symbol $s_{k,n}$ to be decided, however, it is polluted by the effective RIS-empowered cascaded channel. In addition, $I_{2}$ and $I_{3}$ represent the cross-interference terms produced by the noise and the received differential symbol in two time instants, while $I_{4}$ is exclusively produced by the product of the noise in two instants. The symbol decision is performed over the variable $z_{k,n}$ in \eqref{eqn:diff_dec}. Note that the proposed NCDS does not require to obtain the CSI, and hence, the undesirable cascaded channel sounding task can be avoided. To this end, in this paper, we consider any random configuration $\mathbf{\Psi}_{n}$ such that $\psi_{n,m} \sim \eunifc{0}{2\pi}$ for the RIS passive elements, avoiding the overhead to solve any complex optimization problem and then feedback the optimized parameters to the RIS, as in the baseline CDS case. Note that $\mathbf{\Psi_{n}}$ can be randomly set for each OFDM symbol without any restriction (unlike in CDS), however, the continuous configuration of the RIS may unnecessarily increase the energy and/or resource consumption. Hence, it is recommended to update these phase configuration at each data frame (e.g. every $N$ OFDM symbols). It will be shown in the performance evaluation results that the proposed NCDS-based approach provides substantial gains over the baseline CDS, both in terms of computational complexity and achievable performance.

\textit{Remark:} It is reasonable to expect that an optimized RIS configuration design, tailored to the BS-RIS and RIS-UE channels, for the proposed NCDS approach would yield a better decision performance for $s_{k,n}$. However, to obtain this design, forms of CDS will be needed. We leave the investigation for other than random, based on limited or reduced CSI, RIS configurations for the proposed approach for future work.

\subsection{Analysis of the SINR for the IID Channel Model}
According to (\ref{eqn:diff_dec})-(\ref{eqn:diff_dec3}), there are interference and noise terms produced by the differential decoding. The received symbol $z_{k,n}$ should be compared to the transmitted symbol $s_{k,n}$ in order to characterize these undesirable effects, which can be expressed as
\begin{equation} \label{eqn:interference}
\eexpabstwo{s_{k,n}-z_{k,n}}= P_{x}^{2} + \eexpabstwo{z_{k,n}}-2\ereal{\eexp{\eherm{s_{k,n}}z_{k,n}}},
\end{equation}
where the expectation is performed over the subcarriers and OFDM symbols. According to \cite{Victor2018}, the four terms given in (\ref{eqn:diff_dec1}) and (\ref{eqn:diff_dec3}) are statistically independent due to the fact that the channel frequency response, noise, and symbols are independent random variables, and the noise samples between two time instants are also independent. Hence, the two terms in (\ref{eqn:interference}) can be simplified as
\begin{equation} \label{eqn:victor1}
\eexpabstwo{z_{k,n}} = \sum_{i=1}^{4}\eexpabstwo{I_{i}},\,\eexp{\eherm{s_{k,n}}z_{k,n}} = \eexp{\eherm{s_{k,n}}I_{1}},
\end{equation}
and the SINR of the proposed NCDS approach can be defined as
\begin{equation} \label{eqn:sinr}
\rho_{nc} = \frac{P_{x}^{2}}{P_{x}^{2}+\frac{1}{M^{2}B^{2}}\sum_{i=1}^{4}\eexpabstwo{I_{i}}-\frac{2}{MB}\ereal{\eexp{\eherm{s_{k,n}}I_{1}}}}.
\end{equation}

Assuming the IID Rayleigh channel model, each of the expected values in (\ref{eqn:sinr}) can be expressed as (details are included in the Appendix \ref{appendix:iid}):
\begin{equation} \label{eqn:tlast_iid}
	\eexp{\eherm{s_{k,n}}I_{1}} = P_{x}^{2} B\sigma_{h}^{2} M\sigma_{g}^{2},
\end{equation}
\begin{equation} \label{eqn:t1_iid}
	\eexpabstwo{I_{1}} = P_{x}^{2} (1+B)B\sigma_{h}^{4} (1+M)M\sigma_{g}^{4},
\end{equation}
\begin{equation} \label{eqn:t23_iid}
	\eexpabstwo{I_{2}} = \eexpabstwo{I_{3}} = \sigma_{v}^{2} P_{x} B\sigma_{h}^{2} M\sigma_{g}^{2},
\end{equation}
\begin{equation} \label{eqn:t4}
	\eexpabstwo{I_{4}} = B\sigma_{v}^{4}.
\end{equation}
Substituting (\ref{eqn:tlast_iid})-(\ref{eqn:t4}) into (\ref{eqn:sinr}), yields the following expression for the SINR of the proposed NCDS approach:
\begin{equation} \label{eqn:diff_snr_iid}
\begin{split}
\frac{1}{\rho_{nc}^{\text{iid}}} =& 1+\left( 1+\frac{B+M+1}{MB} \right)\sigma_{h}^{4}\sigma_{g}^{4} + \\
& + \left( \frac{\sigma_{v}^{2}}{P_{x}MB}-1\right) 2\sigma_{h}^{2}\sigma_{g}^{2} + \frac{\sigma_{v}^{4}}{P_{x}^{2}M^2 B},
\end{split}
\end{equation}
which indicates that not only the number $B$ of the BS antennas improves the system performance, but also the number $M$ of the RIS passive elements helps to reduce the interference and noise terms. 

For the particular case where $M\rightarrow\infty$ and assuming that the BS is capable of estimating and compensating the average gain of the cascaded channel ($\sigma_{h}^{2}$ and $\sigma_{g}^{2}$), the SINR can be simplified as $\rho_{nc}^{\text{iid}} \approx B$, which corresponds to the SINR expression given in \cite{Ana2015} particularized to a single user and without the noise terms (that vanish thanks to the large number of RIS elements). Note that the average gain of the cascaded channel can be easily estimated by the BS in the random access stage, when the UE transmits the random access sequence to request to enter the system, and hence, no additional reference signal is required. Then, it is compensated by using an automatic gain control.

\subsection{Analysis of the SINR for the Geometric Wideband Channel Model}
Similar to the previous subsection, each of the expected values in (\ref{eqn:sinr}) for the geometric wideband channel model is analytically derived in the Appendix \ref{appendix:geo}, as follows:
\begin{equation} \label{eqn:t1_cdl}
\eexpabstwo{I_{1}} = P_{x}^{2} Q_{4}, \quad \eexp{\eherm{s_{k,n}}I_{1}} = P_{x}^{2} Q_{2},
\end{equation}
\begin{equation} \label{eqn:t23_cdl}
\eexpabstwo{I_{2}} = \eexpabstwo{I_{3}} = P_{x} \sigma_{v}^{2} Q_{2}
\end{equation}
where the terms $Q_{4}$ and $Q_{2}$ are defined as
\begin{equation} \label{eqn:sigma_q4}
\begin{split}
& Q_{4} \triangleq 4 L_{\alpha}^{2} L_{\beta}^{2}  \\
& \times \sum_{c_{\alpha}=1}^{C_{\alpha}} \frac{\sigma_{\alpha_{c}}^{4}}{R_{\alpha}^{2}} \sum_{r_{\alpha}=1}^{R_{\alpha}}
\sum_{c_{\beta}=1}^{C_{\beta}} \frac{\sigma_{\beta_{c}}^{4}}{R_{\beta}^{2}} \sum_{r_{\beta}=1}^{R_{\beta}}
\eexpabsn{\tilde{\mathbf{a}}_{n}\left( c_{\beta},r_{\beta},c_{\alpha}r_{\alpha}\right)}{4},
\end{split}
\end{equation}
\begin{equation} \label{eqn:sigma_q2}
\begin{split}
& Q_{2} \triangleq L_{\alpha} L_{\beta}  \\
&\times \sum_{c_{\alpha}=1}^{C_{\alpha}} \frac{\sigma_{\alpha_{c}}^{2}}{R_{\alpha}} \sum_{r_{\alpha}=1}^{R_{\alpha}}
\sum_{c_{\beta}=1}^{C_{\beta}} \frac{\sigma_{\beta_{c}}^{2}}{R_{\beta}} \sum_{r_{\beta}=1}^{R_{\beta}}
\eexpabstwo{\tilde{\mathbf{a}}_{n}\left( c_{\beta},r_{\beta},c_{\alpha}r_{\alpha}\right) },
\end{split}
\end{equation}
and $\tilde{\mathbf{a}}_{n}\left( c_{\beta},r_{\beta},c_{\alpha}r_{\alpha}\right)$ denotes the joint spatial correlation of the BS and RIS given by (\ref{eqn:eq_corr}).

Comparing (\ref{eqn:t1_cdl})-(\ref{eqn:sigma_q2}) with (\ref{eqn:tlast_iid})-(\ref{eqn:t23_iid}) it turns out that the spatial correlation is upper bounded by the IID case as:
\begin{equation} \label{eqn:comp_q2}
\frac{\eexpabstwo{\tilde{\mathbf{a}}_{n}\left( c_{\beta},r_{\beta},c_{\alpha}r_{\alpha}\right)}}{MB} \leq 1,
\end{equation}
\begin{equation} \label{eqn:comp_q4}
\frac{4\eexpabsn{\tilde{\mathbf{a}}_{n}\left( c_{\beta},r_{\beta},c_{\alpha}r_{\alpha}\right)}{4}}{M^{2}B^{2}} \leq 1 + \frac{B+M+1}{MB}.
\end{equation}
Substituting (\ref{eqn:t1_cdl})-(\ref{eqn:sigma_q2}) in (\ref{eqn:sinr}), the SINR is obtained as
\begin{equation} \label{eqn:diff_snr}
\frac{1}{\rho_{nc}^{\text{geo}}} = 1+ \frac{ Q_{4}}{M^2B^2} + \left( \frac{\sigma_{v}^{2}}{P_{x}MB} - 1\right)\frac{2Q_{2}}{MB} + \frac{ \sigma_{v}^{4} }{P_{x}^{2}M^2B} \leq \frac{1}{\rho_{nc}^{\text{iid}}}.
\end{equation}
Obviously, the spatial correlation of the antennas at the BS and the RIS passive elements is limiting the performance of the system for this geometric wideband channel model as compared to the IID case.

Similarly to the IID case, for the particular case where $M\rightarrow\infty$ and assuming that the BS is again able to estimate and compensate the average gain of the cascaded channel using the random access sequence and an automatic gain control. The SINR can be also simplified as $\rho_{nc}^{\text{geo}} \approx B$.

\section{Error Probability and Complexity Analyses}\label{sec:sercomplex}
In this section, we first present analytical expressions for the SEP performance of the proposed RIS-empowered communication system that is based on NCDS. Then, the complexities of the proposed NCDS system and the baseline CDS, as detailed in Section~III, are discussed.

\subsection{Symbol Error Probability (SEP) Analysis}\label{subsec:sep}
The analysis of an approximated SEP is given in this section to characterize the performance of the proposed NCDS. Reference \cite{Manu2020} has provided an asymptotic analysis for this purpose assuming that the number of the antennas at the BS is very large and assuming only an IID channel model. Therein, the Probability Density Function (PDF) of the decision variable $z_{k,n}$ is approximated as a complex normal distribution. However, this approximation is not very realistic when the number of BS antennas is not large.

Taking into account the analytical expressions for $I_{1}$, $I_{2}$, $I_{3}$, and $I_{4}$ given in (\ref{eqn:diff_dec1}) and (\ref{eqn:diff_dec3}), it holds that the first term depends only on the channel and it is a real random variable, while the rest of the terms depend on the noise and they are complex random variables. Hence, the PDF of $z_{k,n}$ can be approximated as
\begin{equation} \label{eqn:pdf}
f_{z_{k,n}}(u,v) \approx f_{I_{1}}(u) * f_{I_{2},I_{3},I_{4}}(u,v),
\end{equation}
where $f_{I_{1}}(u)$ is the PDF of $I_{1}$, $f_{I_{2},I_{3},I_{4}}(u,v)$ is the joint PDF of $I_{2}$, $I_{3}$ and $I_{4}$, and the variables $u$ and $v$ correspond to the real and imaginary parts of $z_{k,n}$, respectively. As shown in \cite{Manu2020}, the latter joint PDF can be accurately approximated by a zero-mean complex normal distribution with variance
\begin{equation} \label{eqn:pdf_gauss}
\sigma_{s}^{2} = \sum_{i=2}^{4}\eexpabstwo{I_{i}},
\end{equation}
while the PDF expression $f_{I_{1}}(u)$ depends on the chosen propagation channel model. 

For the IID channel model, the scalar product $\eherm{\mathbf{q}_{n-1}^{k}}\mathbf{q}_{k,n}$ is a sum of $B$ statistically independent terms (due to the lack of spatial correlation), where each term's distribution can be accurately approximated by $\egausdr{0}{\eexpabstwo{I_{1}}}$ when $M$ is large enough; this holds from the Central Limit Theorem (CLT) \cite{clt}. Hence,  the PDF of the term $I_{1}$ can be approximated by
\begin{equation} \label{eqn:pdf_real_iid}
f_{I_{1}}(u) \approx \egammd{B}{\eexpabstwo{I_{1}}}.
\end{equation}

On the other hand, for the case of the geometric wideband channel model, there exists spatial correlation among the BS antenna elements and the RIS passive elements. By assuming that the number of clusters/rays is large enough, thus the CLT holds, $f_{I_{1}}(u)$ can be approximated by a zero-mean normal distribution with variance $\eexpabsn{I_{1}}{2}$.
Hence, the SEP of the decision variable $z_{k,n}$ for the $k$-th subcarrier of each $n$-th OFDM symbol, assuming without loss of generality that the transmitted symbol is $s_{k,n}=1$, can be computed from the following double-integral
\begin{equation} \label{eqn:ser}
	P_{e} =  1-\int_{0}^{\infty}\int_{\mathcal{D}_{u}}^{}f_{z_{k,n}}(u,v) \text{d}v \text{d}u, \enspace \mathcal{D}_{u}\in u\tan\left( \frac{\pi}{M_{q}}\right) [-1, 1],
\end{equation}
where $\mathcal{D}_{u}$ denotes the decision region for the particular symbol of interest and $M_{q}$ is the number of the symbols in the PSK constellation. Since (\ref{eqn:ser}) is hard, if not impossible, to be expressed in a closed form, it will be evaluated numerically in the next performance assessment section.

\begin{table}[]
	\centering
	\caption{Complexity Comparison between the Proposed NCDS and the Considered Baseline CDS.}
	\label{tab:complexity}
	\begin{tabular}{|c|c|c|}
		\hline
		& \textbf{\begin{tabular}[c]{@{}c@{}}System Optimization\\ Complexity\end{tabular}} & \textbf{\begin{tabular}[c]{@{}c@{}}Total Number of \\ Complex Products\end{tabular}} \\ \hline
		\textbf{CDS} \cite{Ris06} & $\ecomplexity{R_{t}(B^{3}+M)K}$                                                     & $BK$                                                                           \\ \hline
		\textbf{NCDS} & $-$                                                                                & $(B+1)(K-1)$                                                                   \\ \hline
	\end{tabular}
\end{table}

\subsection{Complexity Analysis}
The complexity evaluation for both CDS and NCDS are summarized in Table \ref{tab:complexity}. The proposed NCDS does not require to solve any complex optimization problem, since we consider a random phase configuration for the RIS. For the differential encoding, the transmitter (i.e., the UE) requires $K-1$ complex products at each OFDM symbol, where the receiver (i.e., BS) needs the same number of complex products for the differential decoding at each RF chain before the symbol decision \cite{Kun2020}, resulting in the total of $B(K-1)$ products.

On the contrary, the CDS not only requires complex products for performing the post-coding, but it also requires to solve the optimization problem \label{eq:maxrate} for the BS combining vector and the RIS phase configuration. Different suboptimal methods can be proposed to avoid the complexity issue, at the expense of decreasing the overall performance. In order to constrain the complexity, the considered baseline CDS implements the iterative method of \cite{Ris06}, whose complexity linearly scales with the number of algorithmic iterations required iterations, $R_{t}$, the number $M$ of RIS passive elements, and the number $K$ of OFDM subcarriers. It is noted that the CDS complexity increases severely with the number of BS antennas (it depends on $B^{3}$), which indicates that a massive MIMO BS is not recommended in RIS-empowered systems based on CDS.

\section{Performance Evaluation Results}\label{sec:num_res}
In this section, several numerical results are provided in order to show the performance of the proposed NCDS, as compared to the considered baseline CDS, and the accuracy of the analytical results. A summary of the simulation parameters is provided in Table \ref{tab:simparam}, the location of each network node is given by the Cartesian coordinates $(x,y,z)$ measured in meters, and $f_{c}$ denotes the carrier frequency. The channel propagation model adopted for the simulation results corresponds to the 3GPP factory scenario of size (60m,120m,3m), with the goal to evaluate the 5G performance \cite{nr-901}. The DS and AS were set the same values for both BS-RIS and RIS-UE channel links. Moreover, we have considered two values for each AS, where the lower values are denoted as the low AS scenario, while the higher values refer to the high AS scenario. Regarding the phase configurations for NCDS, these values are randomly chosen and set to the RIS for each frame ($N$ contiguous OFDM symbols), no matter the coherence time.

\subsection{Verification of the Analysis for the Proposed NCDS}
Figure~\ref{fig:simu_sinr} illustrates the SINR performance as a function of the UE transmit power $P_x$ in dBW of the proposed NCDS with $8$-DPSK for both the IID Rayleigh and the geometric wideband channel models, considering $B=2\times 2$ antennas at the BS and different values $M$ for the number of RIS passive elements. As clearly shown, the performance for the IID channel model corresponds to the best case for all simulated $M$ values. On the other hand, for the particular case of geometric wideband channel, the performance depends on the spatial correlation. When the angular positions of the clusters/rays are separated (i.e., high AS), the performance is better compared to the low AS case. Evidently, his improvement becomes even better for high numbers of $M$. It is also shown in this figure that the SINR analysis given in (\ref{eqn:diff_snr_iid}) and (\ref{eqn:diff_snr}), shown with black solid lines, accurately characterizes the RIS-empowered system performance.

The SEP of $4$-DPSK modulation for the proposed NCDS is demonstrated in Fig.~\ref{fig:simu_ser} for both considered channel models with $B=4\times 4$, high AS, and different values for $M$. The conclusions regarding this performance follow the trends of Fig.~\ref{fig:simu_sinr}. A larger value for $M$ (i.e., an RIS with more passive elements) results in improved performance and the IID channel yields better performance than the geometric one. The approximated SEP provided by Subsection~\ref{subsec:sep} is also plotted in figure, showcasing that the analysis is accurate enough. Furthermore, for the particular case of the IID channel model, the proposed approximation is better than one given in \cite{Manu2020}.
\begin{table}[!t]
	\centering
	\caption{Simulation Parameters}
	\label{tab:simparam}
\begin{tabular}{|c|c|c|c|c|c|}
	\hline
	\textbf{BS location}   & (0,0,3) & \boldsymbol{$f_{c}$}     & 3.5 GHz  & \textbf{ASD} & $7^{\rm o}$, $30^{\rm o}$  \\ \hline
	\textbf{RIS location}  & (3,0,3) & \boldsymbol{$\Delta f$}  & 30 KHz   & \textbf{ASA} & $12^{\rm o}$, $50^{\rm o}$  \\ \hline
	\textbf{UE init. location}   & (6,1,1) & \boldsymbol{$K$}           & 1024     & \textbf{ZSD} & $25^{\rm o}$, $130^{\rm o}$ \\ \hline
	\boldsymbol{$L_{\alpha}$}  & -48 dB  & \boldsymbol{$\sigma_{v}^{2}$}  & -116 dBW & \textbf{ZSA} & $30^{\rm o}$, $150^{\rm o}$ \\ \hline
	\boldsymbol{$L_{\beta}$}   & -59 dB  & \boldsymbol{$N$}  & 140 symb.      & \textbf{DS}  & $0.15$ ms   \\ \hline
\end{tabular}
\end{table}

\subsection{Evaluation of the Efficiency Factor for the Baseline CDS}
The efficiency factor for the considered baseline CDS, as defined in (\ref{eqn:effective_power}), is evaluated considering the parameters given in Table \ref{tab:simparam}, which are taken from the 5G numerology \cite{nr-901}. The CSI of the cascaded channel is obtained hypothetically assuming that the first $M$ OFDM symbols out of $N_{c}$ (coherence time) are exclusively devoted for reference signal transmission. This is a larger overhead than supported in the 5G standard, but, as explained in Section~\ref{sec:baseline}, it is the minimum that allows a CDS-based RIS. In Table \ref{tab:efficiency}, the efficiency factor is numerically evaluated for various values $M$ of the number of the RIS passive elements and UE speeds according to (\ref{eqn:effective_power}). As indicated, an RIS equipped with large numbers of elements can be only applied in scenarios without or with very low mobility, while an RIS with small numbers of elements can be exploited in scenarios with some mobility. It is noted that the exploitation of an RIS with large $M$ values using CDS will also have a negative impact on the system complexity.
\eAddFig{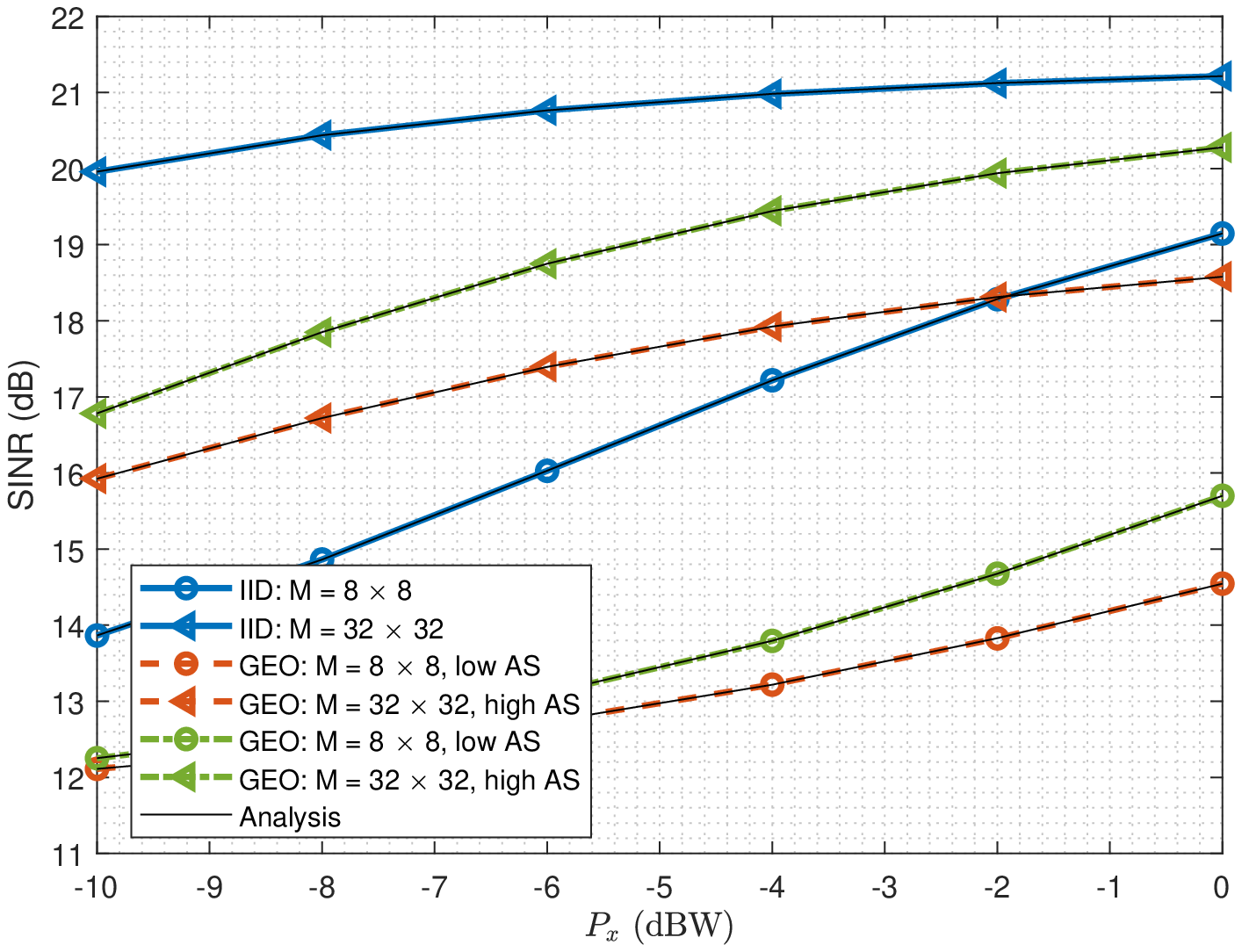}{1}{fig:simu_sinr}{SINR performance in dB of the proposed NCDS for the IID Rayleigh and the geometric wideband channel models for various values $M$ of the RIS passive elements, AS, and for $B=2\times 2$ BS antenna elements.}
\eAddFig{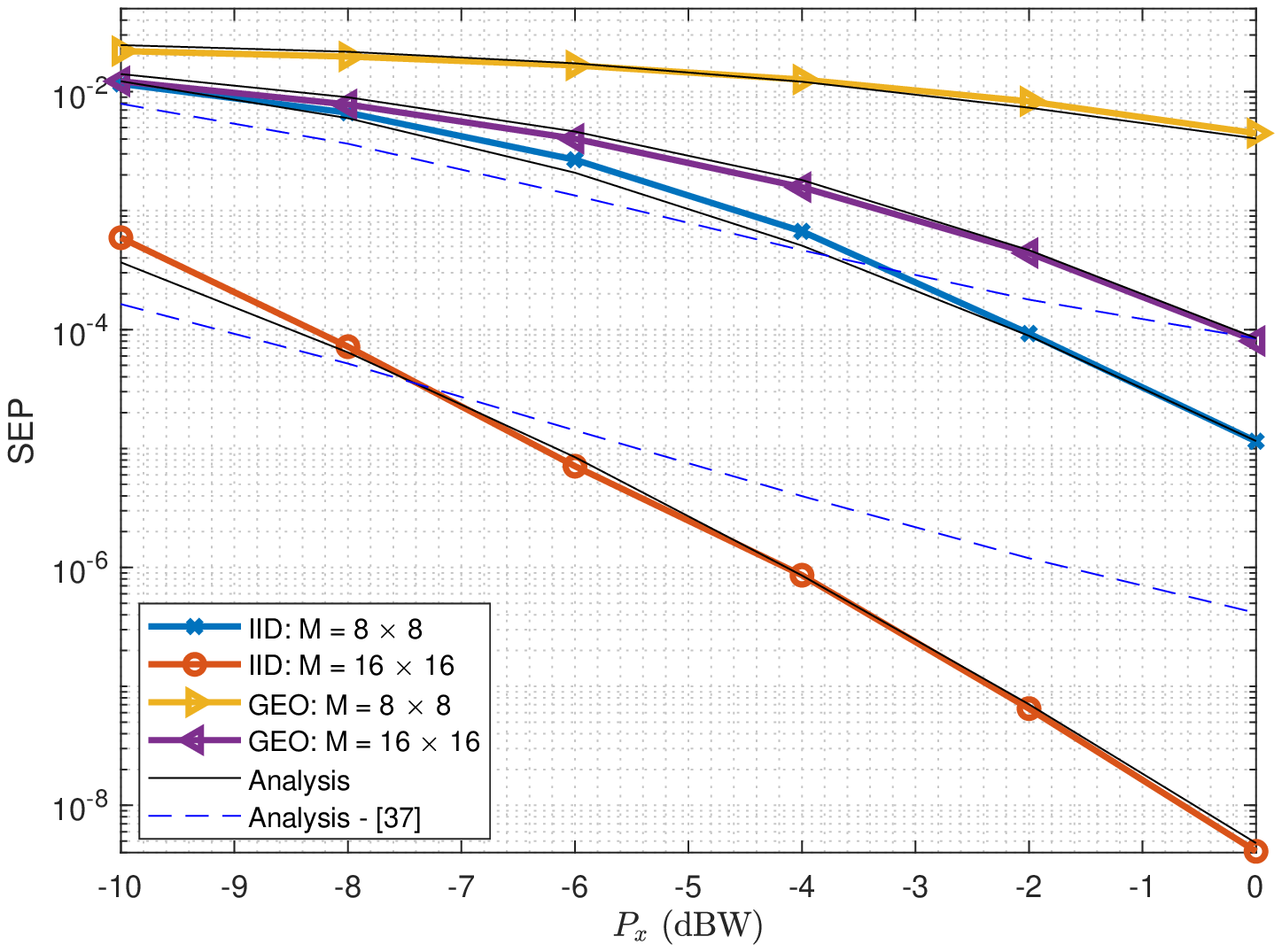}{1}{fig:simu_ser}{SEP performance of the proposed NCDS with $4$-DPSK for the IID Rayleigh and the geometric wideband channel models. Various values $M$ for the RIS passive elements, $B=4\times 4$ BS antennas, as well as high AS (ASD=$30^{\rm o}$, ASA=$50^{\rm o}$, ZSD=$130^{\rm o}$, and ZSA=$150^{\rm o}$) have been considered.}
\begin{table}[!t]
	\centering
	\caption{Efficiency Factor for the Baseline CDS}
	\label{tab:efficiency}
	\begin{tabular}{|c|c|c|c|c|c|}
		\hline
		& \textbf{3 km/h} & \textbf{10 km/h} & \textbf{20 km/h} & \textbf{30 km/h} & \textbf{40 km/h} \\ \hline
		\boldsymbol{$M$}$\mathbf{=16}$ & 0.9738          & 0.9126           & 0.8242           & 0.7377           & 0.6522           \\ \hline
		\boldsymbol{$M$}$\mathbf{=32}$  & 0.9475          & 0.8251           & 0.6484           & 0.4754           & 0.3043           \\ \hline
		\boldsymbol{$M$}$\mathbf{=64}$  & 0.8951          & 0.6503           & 0.2967           & 0                & 0                \\ \hline
		\boldsymbol{$M$}$\mathbf{=128}$ & 0.7902          & 0.3005           & 0                & 0                & 0                \\ \hline
		\boldsymbol{$M$}$\mathbf{=256}$ & 0.5803          & 0                & 0                & 0                & 0                \\ \hline
		\boldsymbol{$M$}$\mathbf{=512}$ & 0.1607          & 0                & 0                & 0                & 0                \\ \hline
		\boldsymbol{$M$}$\mathbf{=1024}$ & 0               & 0                & 0                & 0                & 0                \\ \hline
	\end{tabular}
\end{table}

\subsection{Performance Comparisons between NCDS and CDS}
The SEP performance comparison between the proposed NCDS and the considered baseline CDS for $4$-DPSK and QPSK modulations, respectively, and using a geometric wideband channel with low AS is illustrated in Fig.~\ref{fig:simu_ser_geo_vs}. To perform a fair comparison between the two schemes, we have defined the efficiency factor as follows
\begin{equation} \label{eqn:eff_sel}
	\eta  =
	\left\{\begin{array}{@{}cl}
		1,  & \text{NCDS} \\
		\eta_{c}, & \text{CDS} \\
	\end{array}
	\right..
\end{equation}
This highlights that the proposed NCDS does not suffer any penalization (unlike CDS). Also, it can be deployed at any mobility scenario, exploiting the fact that the differential modulation can be implemented in the frequency dimension, as shown in \cite{Kun2019,Kun2020,Manu2020}. As also concluded from Table~\ref{tab:efficiency}, the proposed NCDS significantly outperforms CDS for large $M$ values for the number of RIS passive elements. Moreover, even for the small $M$ values, the NCDS outperforms CDS, due to the fact that the latter is not able to obtain accurate channel estimates. This happens because the UE's transmit power is in general limited, as also are the amount of resources that can be devoted for CSI estimation while maintaining a reasonable efficiency.

\subsection{Enabling Practical Large RIS with NCDS}
The SEP of the proposed NCDS with $2$-DPSK signaling is depicted in Fig.~\ref{fig:simu_bm_vs} for different values $M$ for the number of RIS elements and $B$ for the the BS antennas, and using a geometric wideband channel model with high AS. It is shown that if any of $B$ and $M$ increases, the NCDS performance improves. It is also evident from the two set of results with the same $MB$ product each that, it is preferable to have a larger $M$ value rather than increasing $B$. Recall that in the proposed NDCS, the RIS phase configurations are randomly chosen, and a higher number $M$ of RIS elements will increase the probability to adequately reflect the signal from the UE to BS. 

In practical RIS implementations \cite{Ris02,huang2018energy}, the phase configurations are constrained to finite sets, since the phase resolution of each RIS element is of the order of a few bits. In Fig.~\ref{fig:simu_bm_vs}, we also consider one-bit quantization of the random phase configuration per data frame, i.e., the phase configurations are randomly obtained from the set of predefined phases ($\psi_{n_{1},m} = \psi_{n_{2},m} \in \mathcal{A}=\left\lbrace0,\pi \right\rbrace$, $1\leq n_{1},n_{2} \leq N$). It can be seen, that this quantization does not affect the NCDS performance, due to the fact that the proposed scheme combines non-coherently all received signals from RIS, and neither channel estimation nor phase configuration optimization are required. 

\eAddFig{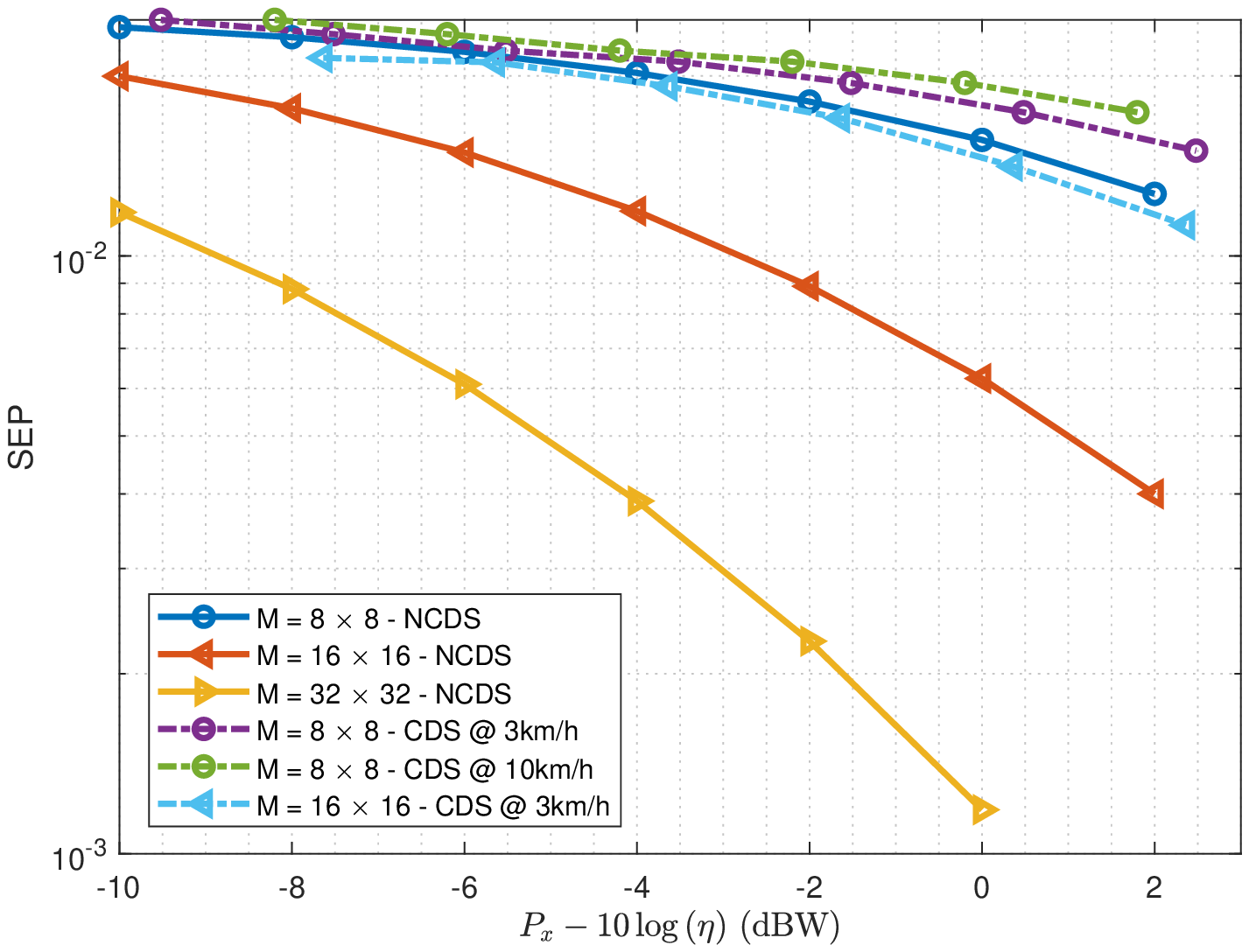}{1}{fig:simu_ser_geo_vs}{SEP performance comparison between the proposed NCDS and the baseline CDS for $4$-DPSK and QPSK, respectively, various numbers $M$ for the RIS elements, $B=2\times 2$ BS antennas, and using a geometric wideband channel model with low AS (ASD=$7^{\rm o}$, ASA=$12^{\rm o}$, ZSD=$25^{\rm o}$, and ZSA=$30^{\rm o}$).}
\eAddFig{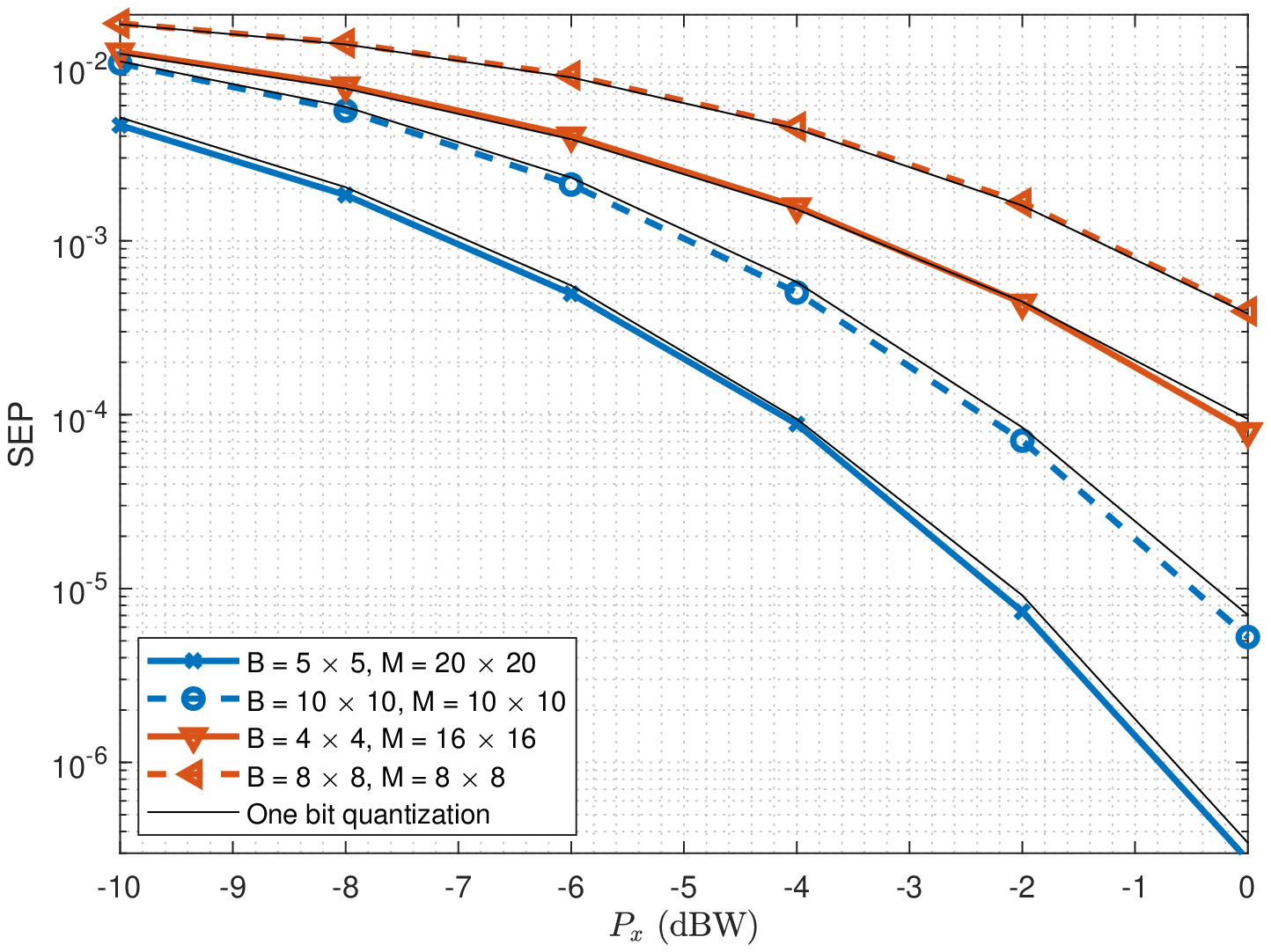}{1}{fig:simu_bm_vs}{SEP performance of the proposed NCDS with $4$-DPSK for different values $M$ for the RIS elements and $B$ for the the BS antennas, using a geometric wideband channel model with high AS (ASD=$30^{\rm o}$, ASA=$50^{\rm o}$, ZSD=$130^{\rm o}$, and ZSA=$150^{\rm o}$). Both random RIS phase configurations and their one-bit quantizations have been simulated.}

\section{Conclusions}\label{sec:conclusion}
This paper investigated NCDS based on differential decoding as an appealing technology for RIS-empowered OFDM wireless communications. The proposed scheme is able to transmit data symbols avoiding any channel training stage, where neither reference signals nor the requirement for solving complex design optimization problems for the BS and RIS parameters are needed. These NCDS will enable the advantages of massive numbers of RIS passive elements, as well as supporting medium/high mobility and/or low-SNR scenarios. 

The proposed method is simple, yet effective, as compared to conventional RIS-empowered systems based on CDS. In contrast to CDS, the presented analysis of the SINR and SEP for both IID Rayleigh and a geometric wideband channel model revealed that the NCDS performance is not only improved by increasing the number of BS antennas, but it can be strongly improved by increasing the number of the RIS passive unit elements. This trend was shown to be present irrespective of the possible low phase resolution of the RIS elements, which happens in practical RIS implementations. For future work, we intend to devise NCDS approaches for RIS-empowered multi-user OFDM communications and study optimized RIS phase configurations, in place of the random ones, that improve further the performance while avoiding time consuming complex optimizations and transmissions of training symbols.

\appendices
\section{Expected Values of the Interference and Noise for the IID Channel Model}\label{appendix:iid}
This appendix provides the derivation of the interference and noise terms required in (\ref{eqn:sinr}) for the considered IID channel model. Assuming that the channel remains quasi-static for two contiguous OFDM symbols and focusing on any $k$-th particular subcarrier, we first use the definitions:
\begin{equation}\label{eq:chan_h_iid}
\mathbf{H}_{k,n}=\mathbf{H}_{n-1}^{k} = \sqrt{L_{\alpha}}\mathbf{A}_{k,n}, \quad 1\leq k \leq K, \quad 2\leq n \leq N,
\end{equation}
\begin{equation}\label{eq:chan_g_iid}
\mathbf{g}_{k,n}=\mathbf{g}_{n-1}^{k} = \sqrt{L_{\beta}}\mathbf{b}_{k,n}, \quad 1\leq k \leq K, \quad 2\leq n \leq N,
\end{equation}
\begin{equation}\label{eq:chan_psi_iid}
\mathbf{\Psi}_{n}=\mathbf{\Psi}_{n-1}, \quad 2\leq n \leq N.
\end{equation}
Note that $\mathbf{\Psi}_{n}$ is omitted in the analysis due to the fact that it is a unit-modulus random variable. Since the channels of the two links (i.e., $\mathbf{H}_{k,n}$ and $\mathbf{g}_{k,n}$) are independent and their corresponding channel coefficients are IID, it holds for any $k$-th subcarrier of each $n$-th OFDM symbol that
\begin{equation}\label{eq:indep1_iid}
\eexp{\eelem{\mathbf{A}_{k,n}}{}{b_{1}m_{1}}\eelem{\mathbf{A}_{k,n}}{}{b_{2}m_{2}}} = 0, \quad b_{1} \neq b_{2}, \enspace \text{or} \enspace m_{1} \neq m_{2},
\end{equation}
\begin{equation}\label{eq:indep2_iid}
\eexp{\eelem{\mathbf{b}_{k,n}}{}{m_{1}}\eelem{\mathbf{b}_{k,n}}{}{m_{2}}} = 0, \quad m_{1} \neq m_{2},
\end{equation}
\begin{equation}\label{eq:indep3_iid}
\eexp{\eelem{\mathbf{A}_{k,n}}{}{bm_{1}}\eelem{\mathbf{b}_{k,n}}{}{m_{2}}} = 0, \quad 1 \leq b \leq B, \quad 1 \leq m_{1},m_{2} \leq M.
\end{equation}

In order to obtain a closed-form expression for $\eexpabstwo{I_{1}}$, we resort to well-known distributions taking into account that the channel coefficients of both links follow a Rayleigh distribution, according to the Subsection~\ref{subsec:iid}. Firstly, focusing on $\mathbf{A}_{k,n}$, it can be easily shown that
\begin{equation}\label{eq:gamma_distri_iid}
\eabsn{\eelem{\mathbf{A}_{k,n}}{}{bm}}{2}\sim\eexpod{\sigma_{\alpha}^{-2}} \rightarrow 
\sum_{b=1}^{B} \eabsn{\eelem{\mathbf{A}_{k,n}}{}{bm}}{2} \sim \egammd{B}{\sigma_{\alpha}^{2}},
\end{equation}
\begin{equation}\label{eq:gamma_mean_iid}
\eexp{\sum_{b=1}^{B} \eabsn{\eelem{\mathbf{A}_{k,n}}{}{bm}}{2}} = B\sigma_{\alpha}^{2}, \quad \evar{\sum_{b=1}^{B} \eabsn{\eelem{\mathbf{A}_{k,n}}{}{bm}}{2}} = B\sigma_{\alpha}^{4},
\end{equation}
\begin{equation}\label{eq:gamma_power_h_iid}
\eexpabstwo{\sum_{b=1}^{B} \eabsn{\eelem{\mathbf{A}_{k,n}}{}{bm}}{2}} = (B+1)B\sigma_{\alpha}^{4}.
\end{equation}
Analogously to $\mathbf{A}_{k,n}$, focusing on $\mathbf{b}_{k,n}$, it can be obtained that
\begin{equation}\label{eq:gamma_power_g_iid}
\eexpabstwo{\sum_{m=1}^{M} \eabsn{\eelem{\mathbf{b}_{k,n}}{}{m}}{2}} = (M+1)M\sigma_{\beta}^{4}.
\end{equation}

Hence, applying (\ref{eq:indep1_iid})-(\ref{eq:gamma_power_g_iid}) and performing some algebraic manipulations, the following simple closed-form expressions can be derived:
\begin{equation}\label{eq:t1_dev_iid}
\begin{split}
\eexpabstwo{I_{1}} & = P_{x}^{2}\eexpabstwo{\sum_{m=1}^{M}\eabsn{\eelem{\mathbf{g}_{k,n}}{}{m}}{2}\sum_{b=1}^{B}\eabsn{\eelem{\mathbf{H}_{k,n}}{}{bm}}{2}} = \\
& = P_{x}^{2}(B+1)B \sigma_{h}^{4} (M+1)M \sigma_{g}^{4},
\end{split}
\end{equation}
\begin{equation}\label{eq:t2_dev_iid}
\begin{split}
\eexpabstwo{I_{2}} & = \eexpabstwo{I_{3}} = \\ 
& = P_{x}\eexpabstwo{\sum_{b=1}^{B}\econj{\eelem{\mathbf{v}}{}{b}}\sum_{m=1}^{M} \eelem{\mathbf{H}_{k,n}}{}{bm} \eelem{\mathbf{g}_{k,n}}{}{m}} = \\
& = P_{x}\sigma_{v}^{2}B\sigma_{h}^{2} M \sigma_{g}^{2}.
\end{split}
\end{equation}
Finally, following the similar procedure as for the derivation of (\ref{eq:t2_dev_iid}), $\eexp{\econj{s_{k,n}}I_{1}}$ is obtained in closed form as
\begin{equation}\label{eq:st1_dev_iid}
	\eexp{\econj{s_{k,n}}I_{1}} = P_{x}^{2}B\sigma_{h}^{2} M \sigma_{g}^{2}.
\end{equation}
Note that the term $\eexpabstwo{I_{4}}$ is obtained using the same procedure as in \cite{Ana2015}, due to the fact that it does not depend on the considered propagation channel model.

\section{Expected Values of the Interference and Noise for the Geometric Wideband Channel Model}\label{appendix:geo}
Similar to the previous appendix, we hereinafter present the derivation of closed-form expression for the interference and noise terms required in (\ref{eqn:sinr}) for the considered geometric wideband channel model. We again assume quasi-static channels between any two contiguous OFDM symbols with $2\leq n \leq N$, and focusing on any $k$-th particular subcarrier we define $\mathbf{\Psi}_{n}$ as in \eqref{eq:chan_psi_iid} as well as $\mathbf{H}_{k,n}$ and $\mathbf{g}_{k,n}$ as follows:
\begin{equation}\label{eq:chan_h}
\mathbf{H}_{k,n}=\mathbf{H}_{n-1}^{k},\,\,\mathbf{g}_{k,n}=\mathbf{g}_{n-1}^{k}.
\end{equation}
Since $\mathbf{H}_{k,n}$ and $\mathbf{g}_{k,n}$ are independent random matrices, their corresponding channel coefficients for all rays and clusters are also independent random variables. This results in:
\begin{equation}\label{eq:indep1}
\eexp{\alpha_{n}^{cr_{1}}\alpha_{n}^{cr_{2}}} = 0, \quad r_{1} \neq r_{2}, \quad 1\leq c \leq C_{\alpha},
\end{equation}
\begin{equation}\label{eq:indep2}
\eexp{\beta_{n}^{cr_{1}}\beta_{n}^{cr_{2}}} = 0, \quad r_{1} \neq r_{2}, \quad 1\leq c \leq C_{\beta},
\end{equation}
\begin{equation}\label{eq:indep3}
\eexp{\alpha_{n}^{c_{1}r_{1}}\alpha_{n}^{c_{2}r_{2}}} = 0, \quad c_{1} \neq c_{2}, \quad 1 \leq r_{1}, r_{2} \leq R_{\alpha},
\end{equation}
\begin{equation}\label{eq:indep4}
\eexp{\beta_{n}^{c_{1}r_{1}}\beta_{n}^{c_{2}r_{2}}} = 0, \quad c_{1} \neq c_{2}, \quad 1 \leq r_{1}, r_{2} \leq R_{\beta},
\end{equation}
\begin{equation}\label{eq:indep5}
\begin{split}
\eexp{\alpha_{n}^{c_{1}r_{1}}\beta_{n}^{c_{2}r_{2}}} & = 0, \quad 1 \leq c_{1} \leq C_{\alpha}, \quad 1 \leq c_{2} \leq C_{\beta},\\
& 1 \leq r_{1} \leq R_{\alpha}, \quad 1 \leq r_{2} \leq R_{\beta}.
\end{split}
\end{equation}

Given (\ref{eq:indep1})-(\ref{eq:indep5}), the interference terms $\eexpabstwo{I_{i}}$ for $1\leq i \leq 3$ as well as $\eexp{\econj{s_{k,n}}I_{1}}$ can be obtained using the superposition principle as
\begin{equation}\label{eq:ti_super}
\eexpabstwo{I_{i}} =
\sum_{c_{\alpha}=1}^{C_{\alpha}}\sum_{r_{\alpha}=1}^{R_{\alpha}}
\sum_{c_{\beta}=1}^{C_{\beta}} \sum_{r_{\beta}=1}^{R_{\beta}}
\eexpabstwo{I_{i}\left( c_{\beta},r_{\beta},c_{\alpha}r_{\alpha}\right)},
\end{equation}
\begin{equation}\label{eq:st1_super}
\eexp{\econj{s_{k,n}}I_{1}} =
\sum_{c_{\alpha}=1}^{C_{\alpha}}\sum_{r_{\alpha}=1}^{R_{\alpha}}
\sum_{c_{\beta}=1}^{C_{\beta}} \sum_{r_{\beta}=1}^{R_{\beta}}
\eexp{\econj{s_{k,n}}I_{1}\left( c_{\beta},r_{\beta},c_{\alpha}r_{\alpha}\right)},
\end{equation}
where $I_{i}\left( c_{\beta},r_{\beta},c_{\alpha}r_{\alpha}\right)$ will be derived in closed form in the sequel. We next capitalize on the fact that the channel coefficients of the BS-RIS and RIS-UE links follow a Rayleigh distribution (see Subsection~\ref{subsec:geo}), hence, for $\alpha_{n}^{cr}$ holds that
\begin{equation}\label{eq:exp_distri}
	\eabsn{\alpha_{n}^{cr}}{2}\sim\eexpod{\frac{R_{\alpha}}{\sigma_{\alpha,c}^{2}}},
\end{equation}
\begin{equation}\label{eq:gamma_mean}
	\eexp{\eabsn{\alpha_{n}^{cr}}{2}} = \frac{\sigma_{\alpha,c}^{2}}{R_{\alpha}}, \quad 
	\evar{\eabsn{\alpha_{n}^{cr}}{2}} = \frac{\sigma_{\alpha,c}^{4}}{R_{\alpha}^{2}},
\end{equation}
\begin{equation}\label{eq:gamma_power_h}
	\eexpabstwo{\eabsn{\alpha_{n}^{cr}}{2}} = 2\frac{\sigma_{\alpha,c}^{4}}{R_{\alpha}^{2}}.
\end{equation}
Analogously to $\alpha_{n}^{cr}$, we can obtained for $\beta_{n}^{cr}$:
\begin{equation}\label{eq:gamma_power_g}
	\eexpabstwo{\eabsn{\beta_{n}^{cr}}{2}} = 2\frac{\sigma_{\beta,c}^{4}}{R_{\beta}^{2}}.
\end{equation}
By using (\ref{eq:exp_distri})-(\ref{eq:gamma_power_g}) and performing some alegraic manipulations, the interference and noise terms produced by each ray and cluster are given by the following closed-form expressions: 
\begin{equation}\label{eq:t1_dev}
\begin{split}
& \eexpabstwo{I_{1}\left( c_{\beta},r_{\beta},c_{\alpha}r_{\alpha}\right)} = P_{x}^{2}L_{\alpha}^{2}L_{\beta}^{2} \eexpabstwo{\eabsn{\alpha_{n}^{c_{\alpha}r_{\alpha}}}{2}} \\ 
& \times  \eexpabstwo{\eabsn{\beta_{n}^{c_{\beta}r_{\beta}}}{2}} \eexpabsn{\tilde{\mathbf{a}}_{n}\left( c_{\beta},r_{\beta},c_{\alpha}r_{\alpha}\right)}{4} = \\
& = 4P_{x}^{2}L_{\alpha}^{2}L_{\beta}^{2} \frac{\sigma_{\alpha,c_{\alpha}}^{4}}{R_{\alpha}^{2}}\frac{\sigma_{\beta,c_{\beta}}^{4}}{R_{\beta}^{2}} \eexpabsn{\tilde{\mathbf{a}}_{n}\left( c_{\beta},r_{\beta},c_{\alpha}r_{\alpha}\right)}{4},
\end{split}
\end{equation}
\begin{equation}\label{eq:t2_dev}
\begin{split}
& \eexpabstwo{I_{2}\left( c_{\beta},r_{\beta},c_{\alpha}r_{\alpha}\right)} = \eexpabstwo{I_{3}\left( c_{\beta},r_{\beta},c_{\alpha}r_{\alpha}\right)} = \\
& = P_{x}L_{\alpha}L_{\beta} \eexpabstwo{\alpha_{n}^{c_{\alpha}r_\alpha}}\eexpabstwo{\beta_{n}^{c_{\beta}r_\beta}}\eexpabstwo{\eelem{\mathbf{v}_{k,n}}{}{b}} \\
& \times \eexpabstwo{\tilde{\mathbf{a}}_{n}\left( c_{\beta},r_{\beta},c_{\alpha}r_{\alpha}\right)} = \\
& = P_{x}\sigma_{v}^{2}L_{\alpha}L_{\beta}\frac{\sigma_{\alpha,c_{\alpha}}^{2}}{R_{\alpha}}\frac{\sigma_{\beta,c_{\beta}}^{2}}{R_{\beta}}\eexpabstwo{\tilde{\mathbf{a}}_{n}\left( c_{\beta},r_{\beta},c_{\alpha}r_{\alpha}\right)},
\end{split}
\end{equation}
where $\tilde{\mathbf{a}}_{n}\left( c_{\beta},r_{\beta},c_{\alpha}r_{\alpha}\right)\in\mathcal{C}^{B\times1}$ represents the joint effect of the spatial correlation of the BS antennas and the RIS passive elements, which is derived as
\begin{equation} \label{eqn:eq_corr}
\begin{split}
&\tilde{\mathbf{a}}_{n}\left( c_{\beta},r_{\beta},c_{\alpha}r_{\alpha}\right) = \mathbf{a}_{\rm BS}\left(\phi_{n}^{c_{\alpha}r_{\alpha}},\theta_{n}^{c_{\alpha}r_{\alpha}}\right)  \\
& \times\mathbf{a}_{\rm RIS}^{H}\left(\bar{\varphi}_{n}^{c_{\alpha}r_{\alpha}},\bar{\vartheta}_{n}^{c_{\alpha}r_{\alpha}}\right)
\mathbf{a}_{\rm RIS}\left(\varphi_{n}^{c_{\beta}r_{\beta}},\vartheta_{n}^{c_{\beta}r_{\beta}}\right).
\end{split}
\end{equation}
Finally, following a similar procedure to the derivation of (\ref{eq:t2_dev}), $\eexp{\econj{s_{k,n}}I_{1}\left( c_{\beta},r_{\beta},c_{\alpha}r_{\alpha}\right)}$ is given by the following closed-form expression:
\begin{equation}\label{eq:st1_dev}
\begin{split}
& \eexp{\econj{s_{k,n}}I_{1}\left( c_{\beta},r_{\beta},c_{\alpha}r_{\alpha}\right)} = P_{x}^{2}L_{\alpha}L_{\beta}\\
& \times\frac{\sigma_{\alpha,c_{\alpha}}^{2}}{R_{\alpha}}\frac{\sigma_{\beta,c_{\beta}}^{2}}{R_{\beta}}\eexpabstwo{\tilde{\mathbf{a}}_{n}\left( c_{\beta},r_{\beta},c_{\alpha}r_{\alpha}\right)}.
\end{split}
\end{equation}

\ifCLASSOPTIONcaptionsoff
  \newpage
\fi
\bibliographystyle{IEEEtran}
\bibliography{./bibtex/IEEEabrv,./bibtex/IEEEexample}{}

\end{document}